\documentclass[aps,prb,fleqn,12pt,amsmath,amssymb,superscriptaddress]{revtex4}
\usepackage{epsfig}
\usepackage{color}
\usepackage{graphicx}
\usepackage{pdfpages}
\usepackage{amssymb,amsmath}
\usepackage{hyperref}

\begin{document}

\title{Effects of the structural distortion on the \\ electronic band structure of {\boldmath $\rm Na Os O_3$} studied within \\ density functional theory and a three-orbital model}
\author{Shubhajyoti Mohapatra}
\affiliation{Department of Physics, Indian Institute of Technology, Kanpur - 208016, India}
\author{Churna Bhandari}
\affiliation{Department of Physics and Astronomy, University of Missouri, Columbia, MO 65211} 
\author{Sashi Satpathy}
\affiliation{Department of Physics and Astronomy, University of Missouri, Columbia, MO 65211} 
\author{Avinash Singh}
\email{avinas@iitk.ac.in}
\affiliation{Department of Physics, Indian Institute of Technology, Kanpur - 208016, India}
\affiliation{Department of Physics and Astronomy, University of Missouri, Columbia, MO 65211} 

\date{\today} 
\begin{abstract}
Effects of the structural distortion associated with the $\rm OsO_6$ octahedral rotation and tilting on the electronic band structure and magnetic anisotropy energy for the $5d^3$ compound NaOsO$_3$ are investigated using the density functional theory (DFT) and within a three-orbital model. Comparison of the essential features of the DFT band structures with the three-orbital model for both the undistorted and distorted structures provides insight into the orbital and directional asymmetry in the electron hopping terms resulting from the structural distortion. The orbital mixing terms obtained in the transformed hopping Hamiltonian resulting from the octahedral rotations are shown to account for the fine features in the DFT band structure. Staggered magnetization and the magnetic character of states near the Fermi energy indicate weak coupling behavior.

%Key words: Spin-orbit coupling, Sodium Osmate, G-type AF State, DFT, Three-Band Model, Electronic Band Structure, Staggered Magnetization 
\end{abstract}
\pacs{75.30.Ds, 71.27.+a, 75.10.Lp, 71.10.Fd}
\maketitle
\newpage

\section{Introduction}

The strongly spin-orbit coupled $5d^3$ osmium compounds $\rm Na Os O_3$ and $\rm Cd_2 Os_2 O_7$ exhibit several novel electronic and magnetic properties. These include continuous metal-insulator transition (MIT) for $\rm Na Os O_3$ that coincides with the antiferromagnetic (AFM) transition ($T_{\rm N} = T_{\rm MIT}$ = 410 K),\cite{shi_PRB_2009} G-type AFM structure with spins oriented along the $c$ axis as seen in neutron and x-ray scattering,\cite{calder_PRL_2012} significantly reduced total (spin $+$ orbital) magnetic moment ($\sim 1 \mu_{\rm B}$) as measured from neutron scattering and ascribed to itinerant-electron behavior due to hybridization between Os $5d$ and O $2p$ orbitals,\cite{calder_PRL_2012} and large spin wave energy gap of 58 meV as seen in resonant inelastic X-ray scattering (RIXS) measurements indicating strong magnetic anisotropy.\cite{calder_PRB_2017} Large spin wave gap has also been observed in the frustrated type I AFM ground state of the double perovskites $\rm Ba_2 Y Os O_6$, $\rm Sr_2 Sc Os O_6$, $\rm Ca_3 Li Os O_6$ in neutron scattering and RIXS studies of the magnetic excitation spectrum,\cite{kermarrec_PRB_2015,taylor_PRB_2016,taylor_PRL_2017} highlighting the importance of spin-orbit coupling (SOC) induced anisotropy despite the nominally orbitally-quenched ions in the $5d^3$ and $4d^3$ systems.

First-principle calculations have been carried out to investigate the electronic and magnetic properties of the orthorhombic perovskite $\rm Na Os O_3$,\cite{du_PRB_2012,jung_PRB_2013} related osmium based perovskites $\rm AOsO_3$ (A=Ca,Sr,Ba),\cite{zahid_JPCS_2015} and double perovskites $\rm Ca_2 Co Os O_6$ and $\rm Ca_2 Ni Os O_6$.\cite{morrow_CM_2016} Density functional theory (DFT) calculations have shown that the magnetic moment is strongly reduced from $3 \mu_{\rm B}$ in the localized-spin picture to nearly $1 \mu_{\rm B}$ (essentially unchanged by SOC) due to itineracy resulting from the strong hybridization of the $t_{2g}$ orbitals with the oxygen $2p$ orbitals, which is significantly affected by the structural distortion.\cite{jung_PRB_2013} Furthermore, from total energy calculations for different spin orientations with SOC included, the easy axis was determined as $\langle 001 \rangle$,\cite{jung_PRB_2013} as also observed by Calder {\em et al.},\cite{calder_PRL_2012} with large energy cost for orientation along the $\langle 010 \rangle$ axis and very small energy difference between orientations along the nearly symmetrical $a$ and $c$ axes. 

A moderate $U$ $\sim 1-2$ eV has been considered in earlier DFT studies for producing the insulating state with a G-type AFM order.\cite{calder_PRL_2012,du_PRB_2012,jung_PRB_2013,kim_PRB_2016} For the distorted structure with SOC, an indirect band gap is seen to open only at a critical interaction strength $U_c \approx 2$ eV,\cite{du_PRB_2012,jung_PRB_2013} where the direct band gap is $\sim$ 0.27 eV.\cite{jung_PRB_2013} Below $U_c$, the indirect band gap becomes negative due to lowering of the bottom of the conduction band. Similarly, metallic band structure is obtained for the undistorted structure, indicating that the distortion-induced band flattening lowers the minimum $U$ required for the AFM insulating state.\cite{jung_PRB_2013} 

Although finite-temperature study of the band structure evolution shows a similar magnitude of $\sim$ 0.2 eV for the direct band gap in the low-temperature limit, extrapolating from the reported low-temperature optical gap of 0.1 eV based on infrared reflectance measurements,\cite{vecchio_SCI_REP_2013} relatively lower renormalized $U \sim 0.7$ eV was obtained, which was ascribed to the SOC-induced reduction of electron mobility resulting from the band flattening in the paramagnetic state, suggesting a weakly correlated regime.\cite{kim_PRB_2016} However, from the reported trend for the band gaps in DFT studies,\cite{jung_PRB_2013} the indirect band gap becomes negative when the direct band gap $\sim$ 0.1 eV, indicating that if structural distortion is included, then $U \sim$ 0.7 eV is probably too low to give an AFM insulating state. In the experimental study also, the energy scale corresponding to the cutoff frequency $\omega_c \approx 15000$ cm$^{-1}$ for conserving the spectral weight was found to be consistent with the range $U \sim 1-2$ eV as considered in the DFT studies.

A recent RIXS study of the electronic and magnetic excitations in NaOsO$_3$ shows that while local electronic excitations do not change appreciably through the MIT, the low energy magnetic excitations present in the insulating state become weakened and damped through the MIT, \cite{vale_arxiv_2017} presumably due to self doping. Such continuous progression towards the itinerant limit through MIT  is suggested to provide physical insight into the nature of the MIT beyond the relativistic Mott or pure Slater type insulators.
 
Besides the Mott-Hubbard and Slater pictures, the ``spin-driven Lifshitz transition" mechanism proposed recently\cite{kim_PRB_2016} is relevant for systems with small electron correlation and large hybridization. This mechanism, which relies on the small indirect band gap in the AFM state in the presence of strong SOC, with the bottom of the upper band descending below the Fermi energy on increasing temperature accompanied with progressively growing small electron pocket, can account not only for the continuous metal-insulator transition but also the concomitant magnetic transition, with low transition temperature despite the large magnetic anisotropy energy. 
Although weak correlation effects are central to the Slater scenario for both $\rm Na Os O_3$ and $\rm Cd_2 Os_2 O_7$ which exhibit continuous MIT concomitant with three dimensional AFM ordering, magnetic interactions and excitations in both compounds have been studied only within the localized spin picture. A minimal three-orbital-model description of the electronic band structure within the $t_{2g}$ sector for the sodium osmate compound $\rm Na Os O_3$ and a microscopic understanding of the large magneto-crystalline anisotropy within such a minimal model have not been investigated so far. 

In this paper, we will therefore investigate the electronic band structure within a minimal 
three-orbital model, specifically aimed at identifying the orbital and directional asymmetry introduced in the electron hopping parameters due to the $\rm Os O_6$ octahedral rotation and tilting. Moderate bandwidth reduction due to the octahedral rotations has been noted as an important contributing factor to the gapped AFM state. In this paper, we will show that this bandwidth reduction is slightly orbitally and directionally asymmetric, which is expected to play a crucial role in the expression of the SOC-induced magnetic anisotropy in this compound. 

The structure of this paper is as follows. DFT investigation of the electronic band structure for $\rm NaOsO_3$ is presented in Sec. II, including discussion of the density of states (DOS) for Os, AFM ordering and reduced magnetic moment, and effect of octahedral rotations on the band structure. In Sections III and IV, the electronic band structure and staggered magnetization are studied within a minimal three-orbital model and compared with DFT results for both the undistorted and distorted structures. The orbital and directional asymmetry in the hopping terms resulting from the cubic symmetry breaking are highlighted in the last part of Sec. III. The SOC-induced magnetic anisotropy is discussed in Sec. V, followed by conclusions in Sec. VI.

\section{Density functional electronic structure}    

Crystal structure of NaOsO$_3$ is orthorhombic with space group $P_{nma}$ (62) consisting of four formula units per unit cell. The experimental lattice constants are $a$=5.384~\AA, $c$=5.328~\AA ~and $b$=7.58 \AA.\cite{shi_PRB_2009} The unit cell is approximately $\sqrt{2}a_0\times \sqrt{2}a_0$ in the $ac$ plane and doubled ($2a_0$) perpendicular to the plane, where $a_0$ is the nearest-neighbor Os-Os distance in the plane. In the plane, the Os octahedra tilt by $12.48^{\circ}$, which is a rotation about the $a$ axis, followed by a second rotation of $ 8.74^{\circ}$ about the $b$ axis as discussed in the Appendix. Along the $b$ axis, a doubling of the unit cell occurs due to the out-of-phase tilting of adjacent Os octahedra. The Wyckoff's positions according to the site symmetry are: Na(4c), Os(4b), O(4c), and O(8d).

The full-potential linearized muffin-tin orbital (LMTO) method \cite{Methfessel,Kotani10,lmsuite} was used to calculate the electronic band structure within the local spin density approximation (LSDA)\cite{vonBarthHedin} for the exchange and correlation functional including the effective on-site Coulomb repulsion $U_{\rm eff}$ (=2 eV) and SOC (LSDA+SO+$U$). In DFT, SOC is a position dependent operator $H_{\rm SOC} = \lambda(r) {\bf L.S}$, where $\lambda(r)= \dfrac{\hbar}{2m^2c^2}\dfrac{1}{r}\dfrac{\partial V}{\partial r}$ in terms of the potential $V$ felt by the electron. On the other hand, in the three-orbital model discussed in the next section, it will be incorporated by the on-site term $\lambda {\bf L.S}$ for every atom. The experimental lattice parameters were used throughout the calculations. The unshifted $\bf k$-space mesh $8\times8\times6$ was used for the Brillouin-zone integration. 

\begin{figure}
\vspace*{0.35cm}
\includegraphics[scale=0.6]{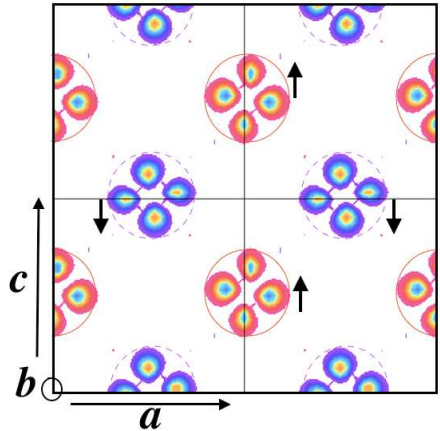}
\caption{(Color Online) Contour plot of the spin density difference $(\rho_\uparrow - \rho_\downarrow)$ on the $ac$ plane calculated for the electron states within 0.2 eV of the conduction band minimum, computed with the LSDA+SO+U method. The outermost contour (dark) represents the spin density difference isosurface level of $4\times 10^{-3}$ e/\AA$^3$, while the innermost (light) represents the spin density difference isosurface level of $3.4\times 10^{-2}$ e/\AA$^3$. The magnetic moments (presented in Sec. III) correspond to the net moment within the muffin-tin spheres indicated by the circles. Arrows indicate the net spins on the Os sites, indicating the AFM structure.}
\label{rho}
\end{figure}

%\begin{figure}
%\includegraphics[width=0.8\linewidth,angle=0]{fig1.eps} %doslapw
%\caption{(Color Online) Density of states (DOS) for NaOsO$_3$ as obtained with the LSDA+SO+U method. Different panels show (a) total DOS in black (red) for spin $\uparrow (\downarrow)$ per unit cell, (b) partial density of states (PDOS) of the $e_g$ states, and (c) exchange splitting in the $t_{2g}$ manifold for a single Os atom.}
%\label{dos}
%\end{figure}

The calculated total and partial DOS for a single Os atom within the LSDA+SO+$U$ method are in agreement with earlier studies. The orbital resolved PDOS shows that states near the Fermi level are mainly $t_{2g}$-like. The magnetic moments are calculated by integrating the spin density within the muffin-tin sphere. In the absence of Coulomb repulsion, electrons are not fully spin polarized and the magnetic moment is therefore relatively small ($0.66 ~\mu_B$). With the inclusion of SOC, it is further reduced due to spin mixing. Finally, when Coulomb interaction is included, the magnetic moment is enhanced, as expected, since the Coulomb term favors spin polarization in order to reduce the Coulomb energy. The calculated (total) magnetic moment  0.88 $\mu_B$ is in reasonable agreement with the experimental value of 1.01 $\mu_B$.\cite{calder_PRL_2012} Note that the calculated value can vary slightly depending on the size of the muffin-tin radius chosen. The calculated magnetic moments using different methods will be presented in Sec. III.

The calculated spin density contours in the $ac$ plane for the electron states in the conduction bands are shown in Fig. \ref{rho}. Each contour represents mainly the $d_{xy}$ part of the $t_{2g}$ orbitals, since the other two orbitals ($d_{xz}, d_{yz}$) have negligible contribution on this plane. The figure clearly shows the G-type AFM ordering. 

Band structures were calculated for both the real crystal structure and the undistorted structure  (with the same experimental lattice constants but without rotations) for the same G-type AFM configuration. We find a net energy gain of about 264 meV (per formula unit) for the distorted structure over the undistorted structure, consistent with previous works.\cite{du_PRB_2012,jung_PRB_2013} The calculated DFT bands with and without structural distortion will be discussed in Sec. III along with the results obtained from the three-orbital model. 

%Here $\Gamma=(0,0,0)$, ${\rm X}=\frac{\pi}{a}(1,0,0)$, ${\rm Y}=\frac{\pi}{c}(0,1,0)$, ${\rm Z}=\frac{\pi}{b}(0,0,1)$, ${\rm S}=(\frac{\pi}{a},\frac{\pi}{c},0)$, ${\rm T}=(0,\frac{\pi}{c},\frac{\pi}{b})$, and ${\rm R}=(\frac{\pi}{a},\frac{\pi}{c},\frac{\pi}{b})$. 

\section{Three-orbital model and electronic band structure}

The electronic and magnetic behaviour of the $5d^3$ compound $\rm Na Os O_3$ involve a complex interplay between SOC, structural distortion, magnetic ordering, Hund's rule coupling, and weak correlation effect. While strong SOC would favor spin-orbital entangled states energetically separated into the $J=1/2$ doublet and $J=3/2$ quartet, strong Hund's rule coupling would favor the spin-disentangled, high-spin nominally $S=3/2$ state in the system with three electrons per Os ion. In the high-spin state, Hund's rule coupling would also effectively enhance the local exchange field, supporting the weak correlation term in the formation of the AFM state in this half-filled system. The enhanced local exchange field would energetically separate the spin up and down states, thus self consistently suppressing the SOC. 

In order to obtain a detailed microscopic understanding of the electronic and magnetic properties including magnetic anisotropy and large spin wave gap induced by spin orbit coupling in this half-filled AFM insulating system, it will be helpful to start with a simplified model for the electronic band structure. In this section, we will therefore consider a minimal three-orbital model involving the $5d$ $yz,xz,xy$ orbitals constituting the $t_{2g}$ sector, aimed at reproducing the essential features of the DFT calculation discussed in the previous section. 

\subsection{Non-magnetic state}
We start with the free part of the Hamiltonian including the local spin-orbit coupling and the band terms represented in the three-orbital basis $(yz\sigma,xz\sigma,xy\bar{\sigma})$\cite{iridate_paper}: 
\begin{equation}
\mathcal{H}_{\rm SO} + \mathcal{H}_{\rm band} = \sum_{{\bf k} \sigma} \psi_{{\bf k} \sigma}^{\dagger} \begin{pmatrix}
{\cal E}_{\bf k} ^{yz} & i \sigma\frac{\lambda}{2} & -\sigma\frac{\lambda}{2} \\
- i \sigma\frac{\lambda}{2} & {\cal E}_{\bf k} ^{xz} & i\frac{\lambda}{2} \\
-\sigma\frac{\lambda}{2} & - i\frac{\lambda}{2} & {\cal E}_{\bf k} ^{xy} \\
\end{pmatrix} \psi_{{\bf k} \sigma} 
\label{three_orb_matrix}
\end{equation} 
where $\lambda$ is the spin-orbit coupling parameter, and ${\cal E}_{\bf k} ^\mu$ are the band energies for the three orbitals $\mu=yz,xz,xy$, defined with respect to a common spin-orbital coordinate system. In the following it will be convenient to distinguish between the band energy contributions from hopping terms connecting opposite sublattices ($\epsilon_{\bf k} ^{\mu}$) and same sublattice (${\epsilon_{\bf k} ^{\mu}}^\prime$). In addition to the orbital-diagonal band terms above, hopping terms involving orbital mixing will be considered below. For simplicity, we have considered the two-sublattice case for illustration, which can be easily extended to the realistic four-sublattice basis considered in the band structure study. 

\subsection{AFM state and staggered field}
Including the symmetry-breaking staggered fields $-s${\boldmath $\sigma . \Delta_\mu$} for the three orbitals $\mu=yz,xz,xy$, where $s=\pm 1$ for the two sublattices A/B, the staggered-field contribution: 
\begin{equation}
\mathcal{H}_{\rm sf} = \sum_{{\bf k} \sigma s} 
s\sigma \psi_{{\bf k} \sigma s}^{\dagger} 
\begin{pmatrix}
-\Delta_{yz}^z & 0 & 0 \\
0 & -\Delta_{xz}^z & 0 \\
0 & 0 & +\Delta_{xy}^z \\
\end{pmatrix} \psi_{{\bf k} \sigma s} 
\label{stag_field_matrix}
\end{equation} 
for ordering in the $z$ direction. For general ordering direction with components {\boldmath $\Delta_\mu$}= $(\Delta_\mu^x,\Delta_\mu^y,\Delta_\mu^z)$ for orbital $\mu$, the spin-space representation of the staggered field contribution:  
\begin{equation}
\mathcal{H}_{\rm sf} = 
\sum_{{\bf k} \sigma \sigma' s \mu}  \psi_{{\bf k} \sigma s \mu}^{\dagger} 
\begin{pmatrix} -s \makebox{\boldmath $\sigma . \Delta_\mu$}
\end{pmatrix}_{\sigma \sigma'} \psi_{{\bf k} \sigma' s \mu} 
= \sum_{{\bf k} \sigma \sigma' s \mu} s \psi_{{\bf k} \sigma s \mu}^{\dagger} 
\begin{pmatrix} -\Delta_\mu^z & -\Delta_\mu^x + i \Delta_\mu^y \\
-\Delta_\mu^x - i \Delta_\mu^y & \Delta_\mu^z  \\
\end{pmatrix}_{\sigma \sigma'} \psi_{{\bf k} \sigma' s \mu} 
\end{equation} 

Combining the SO, band, and staggered field terms, the total Hamiltonian is given below in the composite three-orbital, two-sublattice basis, showing the hopping terms connecting same and opposite sublattices, and the staggered field contribution (for $z$ direction ordering). Also included are the hopping terms involving orbital mixing between $yz$, $xz$ and $xy$ orbitals due to the structural distortion resulting from the octahedral rotation and tilting. Involving nearest-neighbor (NN) hopping, these orbital mixing terms are placed in the sublattice-off-diagonal $(s\bar{s})$ part of the Hamiltonian:
\begin{eqnarray}
\mathcal{H}_{\rm SO} + \mathcal{H}_{\rm band} + \mathcal{H}_{\rm sf} 
& = & \sum_{{\bf k} \sigma s} \psi_{{\bf k} \sigma s}^{\dagger} \left [ \begin{pmatrix}
{\epsilon_{\bf k} ^{yz}}^\prime & i \sigma\frac{\lambda}{2} & -\sigma\frac{\lambda}{2} \\
- i \sigma\frac{\lambda}{2} & {\epsilon_{\bf k} ^{xz}}^\prime & i\frac{\lambda}{2} \\
-\sigma\frac{\lambda}{2} & - i\frac{\lambda}{2} & {\epsilon_{\bf k} ^{xy}}^\prime \end{pmatrix} 
-s\sigma \begin{pmatrix}
\Delta_{yz}^z & 0 & 0 \\
0 & \Delta_{xz}^z & 0 \\
0 & 0 & -\Delta_{xy}^z \\
\end{pmatrix} \right ]
\psi_{{\bf k} \sigma s} \nonumber \\
& + & 
\sum_{{\bf k} \sigma s} \psi_{{\bf k} \sigma s}^{\dagger}
\begin{pmatrix}
\epsilon_{\bf k} ^{yz} & \epsilon_{\bf k} ^{yz|xz} & \epsilon_{\bf k} ^{yz|xy} \\
-\epsilon_{\bf k} ^{yz|xz} & \epsilon_{\bf k} ^{xz} & \epsilon_{\bf k} ^{xz|xy} \\
-\epsilon_{\bf k} ^{yz|xy} & -\epsilon_{\bf k} ^{xz|xy} & \epsilon_{\bf k} ^{xy} \end{pmatrix} 
\psi_{{\bf k} \sigma \bar{s}} 
\label{three_orb_two_sub}
\end{eqnarray} 

For straight Os-O-Os bonds (undistorted structure), all hopping terms between NN Os ions are orbital-diagonal with no orbital mixing. Due to twisting of Os-O-Os bonds associated with rotation and tilting of the $\rm Os O_6$ octahedra in $\rm Na Os O_3$, local axes of $\rm OsO_6$ octahedra are alternatively rotated, giving rise to mixing between orbitals.

The staggered fields {\boldmath $\Delta_\mu$} are self-consistently determined from:
\begin{equation}
2\makebox{\boldmath $\Delta_\mu$} = U_\mu \makebox{\boldmath $m_\mu$} + 
J_{\rm H} \sum_{\nu \neq \mu} \makebox{\boldmath $m_\nu$}
\label{selfcon}
\end{equation}
in terms of the staggered magnetizations {\boldmath $m_\mu$}=($m_\mu ^x,m_\mu ^y,m_\mu ^z$) for the three orbitals $\mu$. The above staggered fields arise from the Hartree-Fock (HF) approximation of the electron correlation terms: $\sum_{i\mu} U_\mu n_{i\mu\uparrow} n_{i\mu\downarrow} - 2J_{\rm H} \sum_{i,\mu \ne \nu} {\bf S}_{i\mu} . {\bf S}_{i\nu}$ in the AFM state. For ordering in the $z$ direction ({\boldmath $\Delta_\mu$}=$\Delta \hat{z}$), the staggered magnetizations are evaluated from the spin-dependent electronic densities $n_{\mu\sigma}$ obtained by summing over the occupied states: 
\begin{equation}
m_\mu ^z (\Delta) = [n_{\mu\uparrow}^A - n_{\mu\downarrow}^A](\Delta) = [n_{\mu\uparrow}^A - n_{\mu\uparrow}^B](\Delta) = \frac{1}{N} \sum_{{\bf k},l} ^{E_{{\bf k}l\mu}^\sigma < E_{\rm F}} [|\phi_{{\bf k}l\uparrow}^\mu|^2 - |\phi_{{\bf k}l\downarrow}^\mu|^2]_A (\Delta)
\label{magneqn}
\end{equation}
where $l$ is the branch label and $N$ is the total number of ${\bf k}$ states. In practice, it is easier to consider a given $\Delta$ and self-consistently determine the interaction strength $U_\mu$ from Eq. \ref{selfcon}. We will also consider the energy resolved staggered magnetization $\delta m = (1/3N_k) \sum_{{\bf k}l\mu} ' (n_{{\bf k}l\mu}^\uparrow - n_{{\bf k}l\mu}^\downarrow)$ averaged over the three orbitals, where $\sum_{k}'$ indicates that band energies $E_{{\bf k}l\mu}$ lie in a narrow energy range $E \pm \delta E_{\rm band}/2$ of width $\delta E_{\rm band}$.  

The orbital moment was calculated using the AFM band eigenstates $| {\bf k} \rangle $ as below:
\begin{eqnarray}
m_{\rm orb}^z & = & \sum_{{\bf k},\sigma} \left [ 
| \langle \sigma,m_L=+1 | {\bf k} \rangle |^2  -
| \langle \sigma,m_L=-1 | {\bf k} \rangle |^2 \right ] \nonumber \\
& = & \sum_{{\bf k},l,\sigma} \left [ 
| \phi_{{\bf k}l\sigma} ^{yz} - i  \phi_{{\bf k}l\sigma} ^{xz}  |^2  -
| \phi_{{\bf k}l\sigma} ^{yz} + i  \phi_{{\bf k}l\sigma} ^{xz}  |^2 
\right ]
\label{mag_orb}
\end{eqnarray}
where $|m_L = \pm 1, \sigma \rangle = (|yz,\sigma \rangle \pm i |xz,\sigma \rangle )/\sqrt{2}$ are the base states of the orbital angular momentum operator $L_z$ with eigenvalues $m_L = \pm 1$.  

\begin{figure}
\vspace*{0mm}
\hspace*{0mm}
\psfig{figure=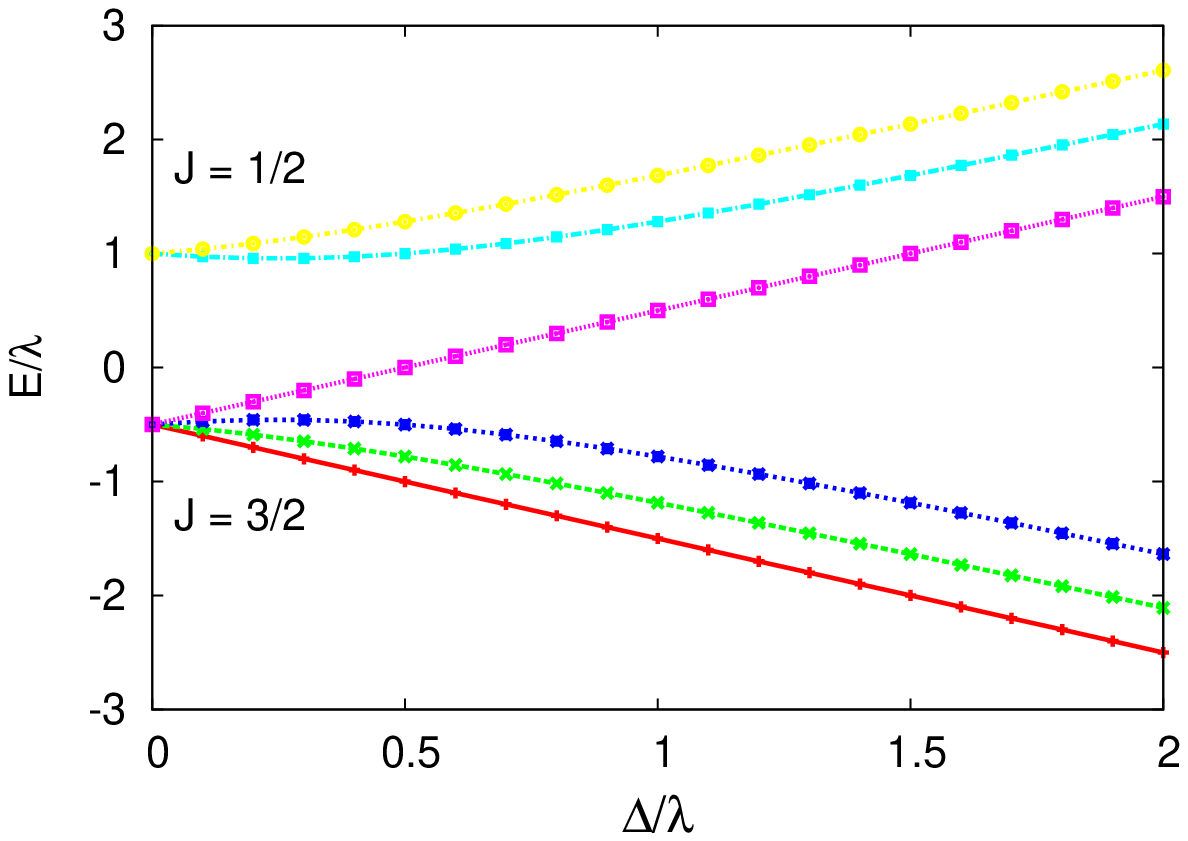,angle=0,width=80mm} 
\psfig{figure=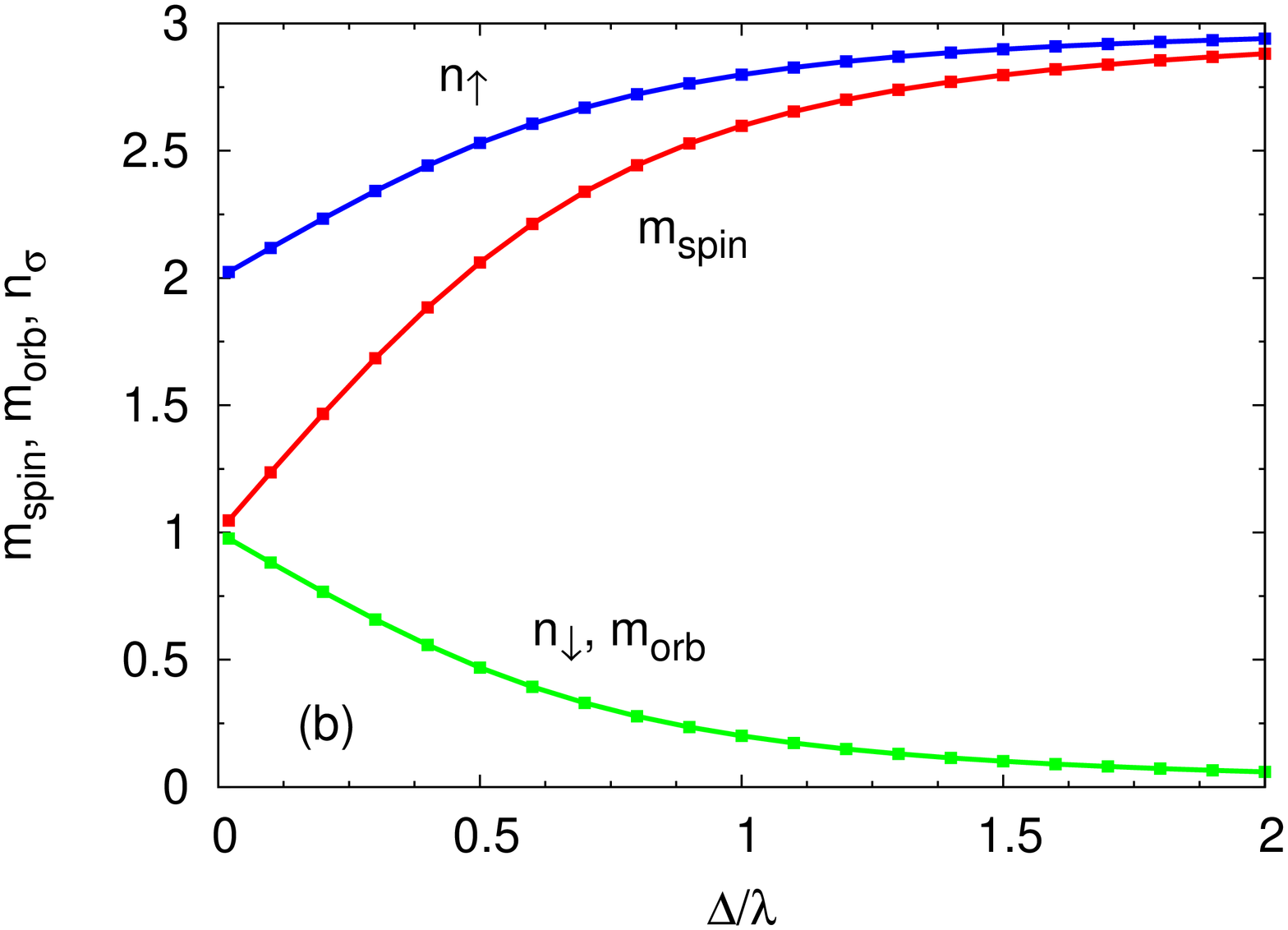,origin=br,angle=-90,width=80mm} 
\caption{Evolution of (a) the SOC-split energy levels and (b) spin and orbital moments along with  electron densities in the atomic limit with increasing exchange field.} 
\label{atomic}
\end{figure}

It is instructive to start with the atomic limit and consider how the SOC-split energy levels (the J=1/2 doublet and J=3/2 quartet) evolve with increasing local exchange field {\boldmath $\Delta$} (assumed identical for all three orbitals for simplicity). While SOC tends to entangle the states with respect to orbital and spin, the exchange field tends to energetically disentangle the spin up and down states for all orbitals. The resulting competition is shown in Fig. \ref{atomic}(a). The states which evolve linearly in energy with increasing $\Delta$ are the $J=3/2$ sector states $(|yz,\sigma \rangle + |xz,\sigma \rangle )/\sqrt{2}$ with no spin mixing. Effect of this disentanglement is also evident from the growth of the local moment with staggered field $\Delta$ as shown in Fig. \ref{atomic}(b).
 
\subsection{Tight-binding model for {\boldmath $\rm Na Os O_3$}} 
This orthorhombic-structure compound has four Os atoms per unit cell. This is because while neighboring $\rm OsO_6$ octahedra within the $a-c$ plane undergo staggered rotation with respect to $b$ axis by angle $\sim 9^\circ$, the octahedral rotations are same for $b$-direction neighbors, which does not conform with the staggered (G-type) magnetic order. The $\rm OsO_6$ octahedra also undergo staggered tilting by angle $\sim 12^\circ$ about the $a$ axis.

For the undistorted case, we will consider identical hopping terms for all three orbitals and for equivalent directions corresponding to the cubic symmetry. Here $t_1$ ($\pi$ overlap), $t_2$ ($\sigma$ overlap), and $t_3$ ($\pi$ overlap) are the first, second, and third neighbor hopping terms, respectively. Furthermore, $t_{1\delta}$ is the first neighbor hopping term corresponding to the $\delta$ overlap, again for all three orbitals. An energy offset $\epsilon_{xy}$ for the $xy$ orbital relative to the degenerate $yz/xz$ orbitals has been included to allow for any tetragonal splitting. The various band dispersion terms in Eq. (\ref{three_orb_two_sub}) are given by: 
\begin{eqnarray}
\varepsilon^{xy}_{\bf k} &=& -4t_1 \cos{(k_x/2)} \cos{(k_y/2)} -2t_{1\delta} \cos k_z
\nonumber \\
{\varepsilon^{xy}_{\bf k}}^{\prime} &=& - 2t_{2} (\cos{k_x} + \cos{k_y}) -
4t_{3}\cos{k_x}\cos{k_y} + \epsilon_{xy} \nonumber \\
\varepsilon^{yz}_{\bf k} &=& -2 [ t_1 \cos{\{(k_x - k_y)/2\}} + t_1 \cos{k_z} + t_{1\delta}
\cos{\{(k_x + k_y)/2\}} ] \nonumber \\
{\varepsilon^{yz}_{\bf k}}^{\prime} &=& -4 t_{2} \cos{\{(k_x - k_y)/2\}} \cos{k_z} - 2t_{3}[\cos(k_x - k_y) + \cos(2k_z)] \nonumber \\
\varepsilon^{xz}_{\bf k} &=& -2 [t_1 \cos{\{(k_x + k_y)/2\}} + t_1 \cos{k_z} + t_{1\delta} \cos{\{(k_x - k_y)/2\}} ] \nonumber \\
{\varepsilon^{xz}_{\bf k}}^{\prime} &=& -4 t_{2} \cos{\{(k_x + k_y)/2\}} \cos{k_z} - 2t_{3}[\cos(k_x + k_y) + \cos(2k_z)]  \nonumber \\
\varepsilon^{yz|xz}_{\bf k} &=& -4t_{m1} \cos{(k_x/2)} \cos{(k_y/2)}  \nonumber \\
\varepsilon^{yz|xy}_{\bf k} &=& + 2 t_{m2} [ \cos{\{(k_x + k_y)/2\}} + 2\cos{\{(k_x -
k_y)/2\}} + \cos{k_z} ] \nonumber \\
\varepsilon^{xz|xy}_{\bf k} &=& -2 t_{m2} [ 2\cos{\{(k_x + k_y)/2\}} + \cos{\{(k_x -
k_y)/2\}} + \cos{k_z} ] 
\label{3_orb_model}
\end{eqnarray}

The ${\bf k}$-space directions correspond to the crystal axes, whereas the $t_{2g}$ sector orbitals ($xz,yz,xy$) and the spin space directions are defined with respect to the ${\rm OsO_6}$ octahedra with coordinate axes along the ${\rm Os-O}$ directions as in Fig. \ref{fig-cube} of Appendix. Here $k_x$ and $k_y$ are in unit of $(\sqrt{2} a_0)^{-1}$ while $k_z$ is in unit of $(2 a_0)^{-1}$ corresponding to the $\sqrt{2}a_0 \times \sqrt{2}a_0 \times 2a_0$ unit cell.

Mixing between $xz$ and $yz$ orbitals and between $xy$ and $xz,yz$ orbitals is represented by the first neighbor hopping terms $t_{m1}$ and $t_{m2}$, respectively. The orbital mixing hopping terms are related to the $\rm Os O_6$ octahedral rotation and tilting angles through $t_{m1}=V_{\pi} \theta_r = t_1 \theta_r$ and $t_{m2}=V_{\pi} \theta_t /\sqrt{2} = t_1 \theta_t /\sqrt{2}$ in the small angle approximation. This follows from the transformation of the hopping Hamiltonian matrix in the rotated basis, as shown in Eq. \ref{eqna8} of the Appendix. The $\sqrt{2}$ factor above accounts for the resolution of the tilt angle about the crystal $a$ axis into the two Os-O-Os axes oriented at angle $\pi/4$. Both the orbital mixing hopping terms will be set to 0.15 in units of $t_1$, corresponding to the tilting and rotation angles of approximately 0.2 and 0.15 radian (12 and 9 degrees). The orbital mixing hopping terms have the usual antisymmetry properties: $[\varepsilon^{yz|xz}_{\bf k}]_{ss'} = -[\varepsilon^{xz|yz}_{\bf k}]_{ss'} = [\varepsilon^{yz|xz}_{\bf k}]_{s's}$ etc. For the undistorted structure, we will set the orbital mixing terms to zero.

\begin{table}[!]
\centering
\caption{Hopping parameter values in the three-orbital-model for the undistorted structure in terms of the energy scale unit $|t_{1}|$.}
\label{table-1}
\begin{ruledtabular}
\begin{tabular}{l c  c  c  c  c   c  c  c}
$t_{1}$ & $t_{2}$ & $t_{3}$ & $t_{1\delta}$ & $t_{m1}$ & $t_{m2}$ & $\epsilon_{xy}$ \\ 
\hline 
-1.0 & 0.3 & 0 & 0 & 0 & 0 & 0 \\
\end{tabular} 
\end{ruledtabular}
\end{table}

\begin{figure}
\vspace*{0mm}
\hspace*{0mm}
\psfig{figure= 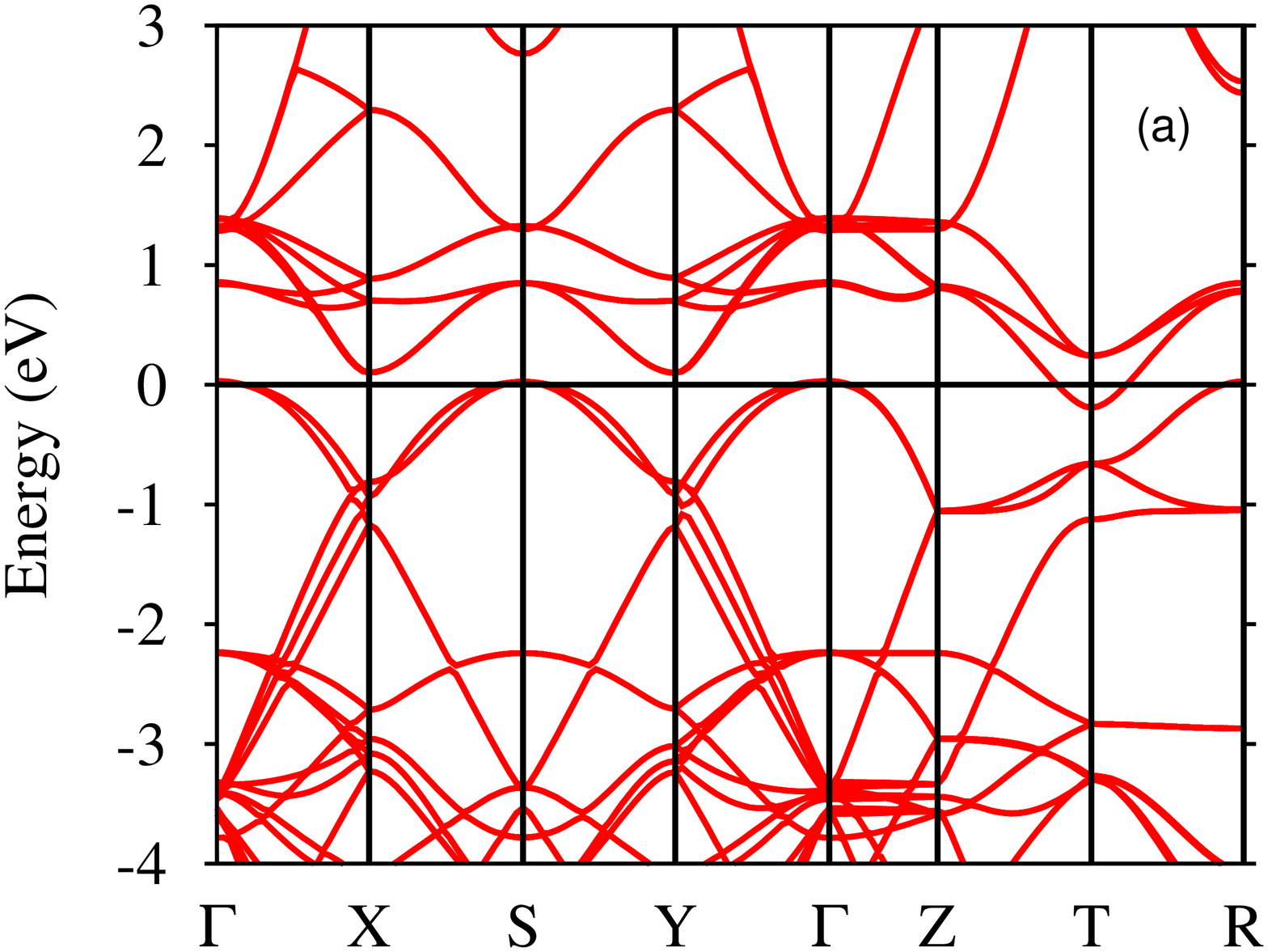,angle=-90,width=80mm}  
\psfig{figure= 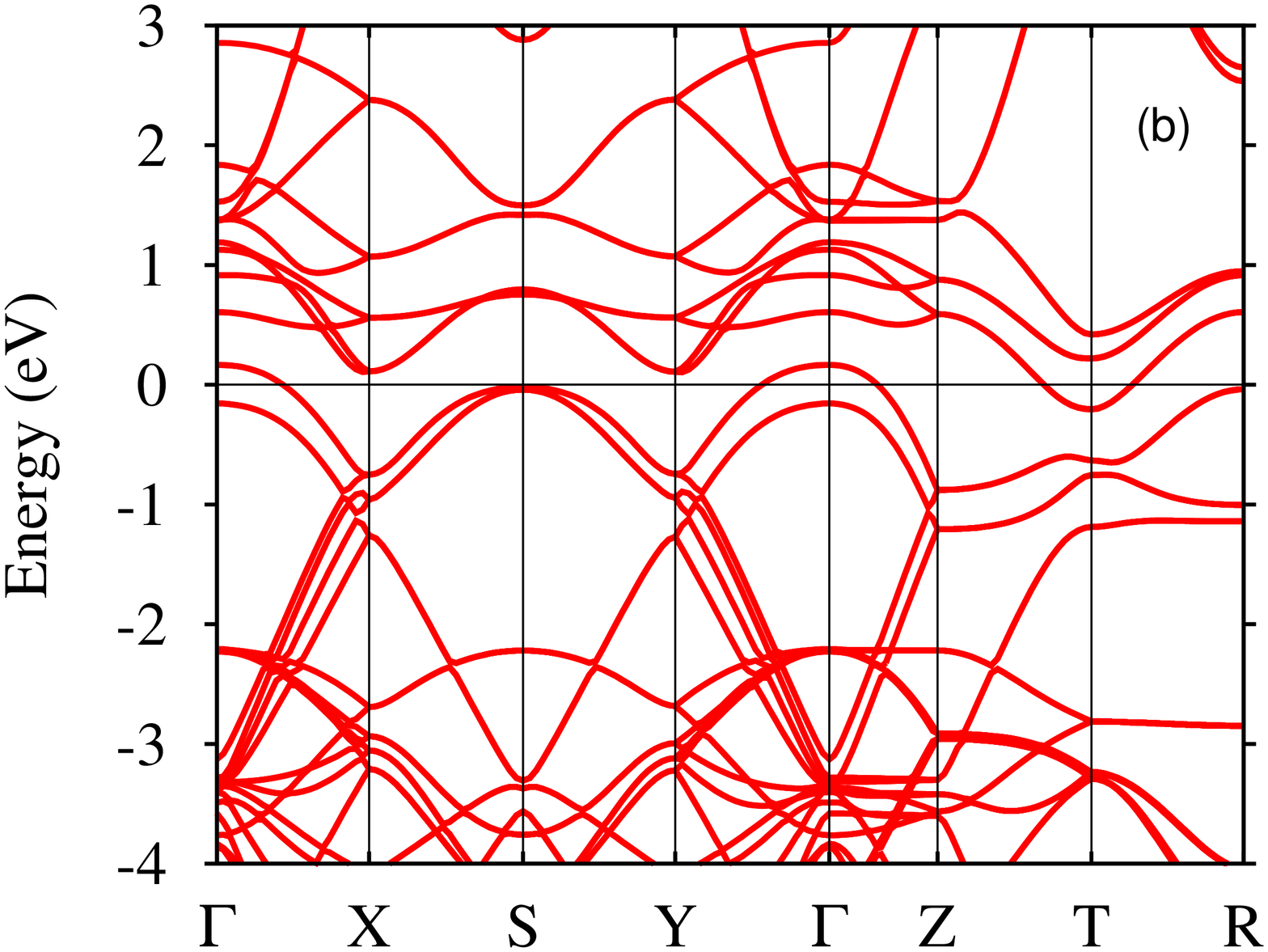,angle=-90,width=80mm} 

\vspace{0.5cm}

\psfig{figure=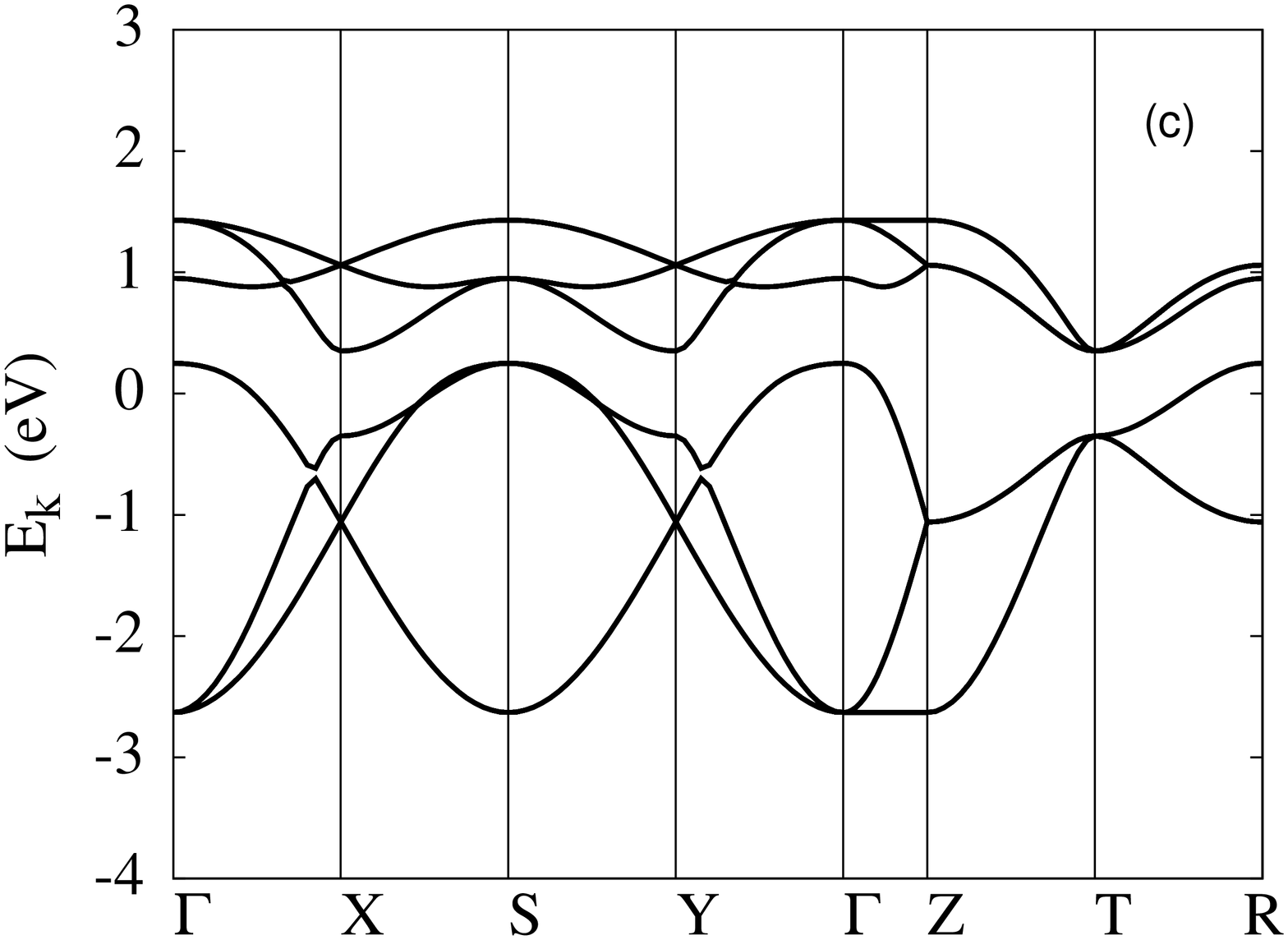,angle=-90,width=80mm}  
\psfig{figure=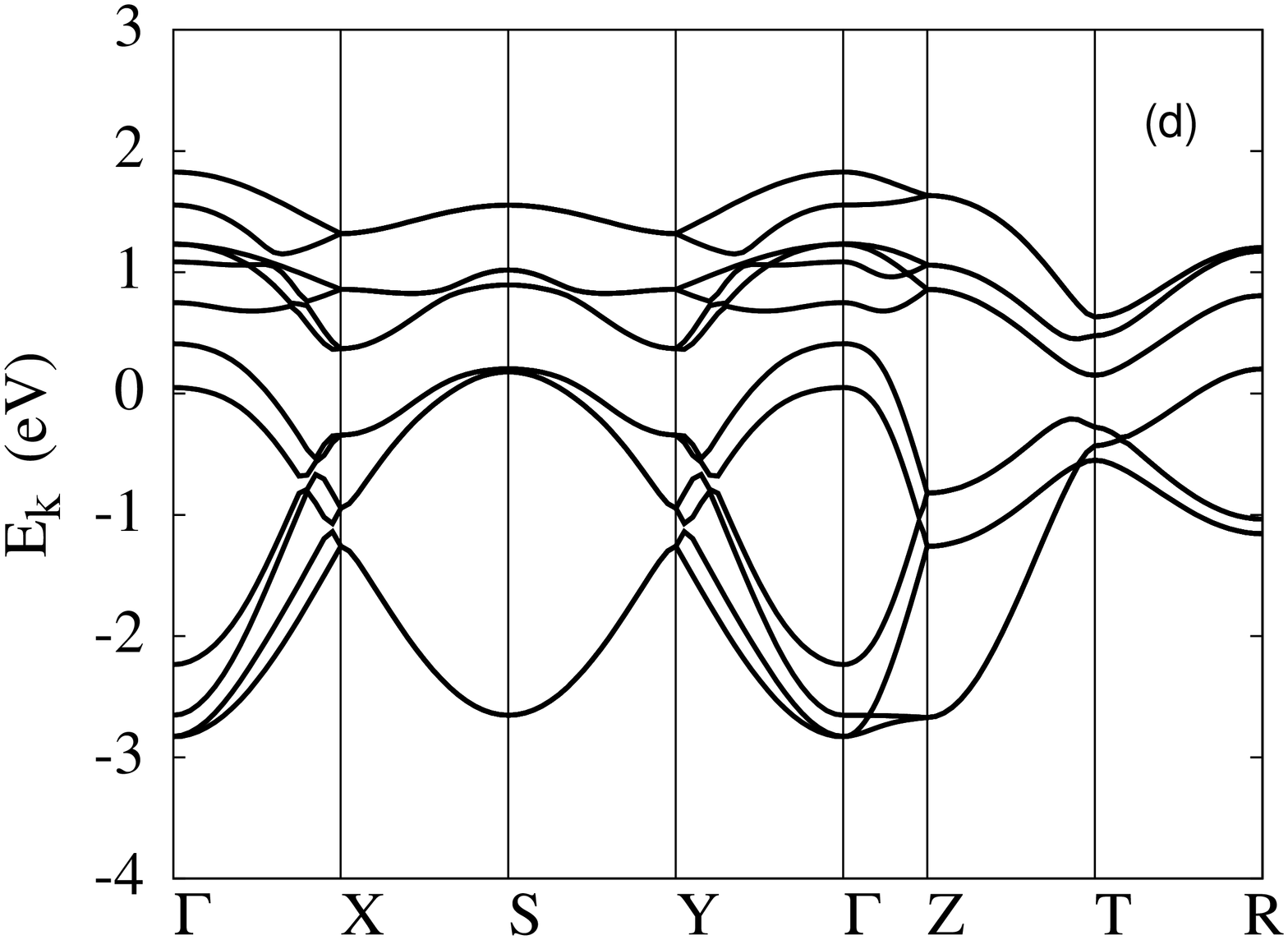,angle=-90,width=80mm}  
\caption{Electronic band structures in the AFM state obtained for the undistorted structure from DFT calculation (a,b), and from the three-orbital model (c,d), without (a,c) and with (b,d) SOC. The momentum-space path considered is: $\Gamma (0,0,0) \rightarrow X (\pi,0,0)  \rightarrow S (\pi,\pi,0)  \rightarrow Y (0,\pi,0)  \rightarrow \Gamma (0,0,0)  \rightarrow Z (0,0,\pi) \rightarrow T (0,\pi,\pi) \rightarrow R (\pi,\pi,\pi)$.} 
\label{band_undis}
\end{figure}

Figure \ref{band_undis} shows the calculated band structures for the undistorted case as obtained from DFT calculation without SOC (a) and with SOC (b), and correspondingly from the three-orbital model (c,d). Band energies are shown along high symmetry directions in the Brillouin zone. To reproduce the essential features of the DFT band structure, we have taken hopping-parameter values as listed in Table \ref{table-1}, staggered field $\Delta = 0.7$, and $\lambda = 0.8$, with energy scale $|{t_1}| = 500$ meV ($\Delta=0.35$ eV and $\lambda = 0.4$ eV). The separation into two groups of three bands above and three below the Fermi energy corresponds to the scenario where Hund's rule coupling dominates over spin-orbit coupling in the atomic limit. For no SOC, the upper band minima are degenerate at $X,Y,T$, but the minimum at $T$ becomes lower when SOC is included. As seen from Fig. \ref{band_undis}(d), strong SOC-induced splitting of the energy bands (e.g., around $\Gamma$) results in negative indirect band gap. The electronic band structure calculated from the three-orbital model [Fig. \ref{band_undis}(c,d)] is broadly consistent with DFT results [Fig. \ref{band_undis}(a,b)]. In general, the strength of SOC in Osmates is about 0.3 eV - 0.4 eV, so the SOC value taken above is consistent with this range.

%realistic value (0.5 eV) reported for $5d$ electrons}.\cite{du_PRB_2012}

For the distorted structure, we will allow for orbital and directional asymmetry in the hopping terms corresponding to the cubic symmetry breaking. In the following, $t_{n(s)} ^{(\mu)}$ and $t_{n(d)} ^{(\mu)}$ refer to the $n^{\rm th}$ neighbor hopping term for orbital $\mu$ connecting sites in the same $(s)$ and different $(d)$ $xy$ planes, respectively. The $xy$ planes here are identified with respect to the octahedral rotation axis (taken as $z$). Due to the lifting of the cubic symmetry in the distorted case, and the resulting four-sublattice structure (A,B and A$'$,B$'$ in alternating $xy$ planes) of the Os lattice, first neighbors in the same $xy$ plane involve AB / A$'$B$'$, whereas second neighbors involve AA / BB / A$'$A$'$ / B$'$B$'$. However, first neighbors in different $xy$ planes involve AB$'$ / A$'$B, whereas second neighbors involve AA$'$ / BB$'$. 

It should be noted that for the $xy$ orbital, both first and second neighbor hopping terms connect sites in the same layer only. Similarly, for the $xz,yz$ orbitals, the second neighbor hopping terms connect sites in different layers only. However, the first neighbor hopping terms for the $xz,yz$ orbitals connect sites in both same and different layers. Therefore, in the following, we will consider $t_{1(s)} ^{(xy)} \ne t_{1(d)} ^{(xz)} = t_{1(d)} ^{(yz)}$ and $t_{2(s)} ^{(xy)} \ne t_{2(d)} ^{(xz)} = t_{2(d)} ^{(yz)}$ etc. 

\begin{figure}
\vspace*{0mm}
\hspace*{0mm}
\psfig{figure=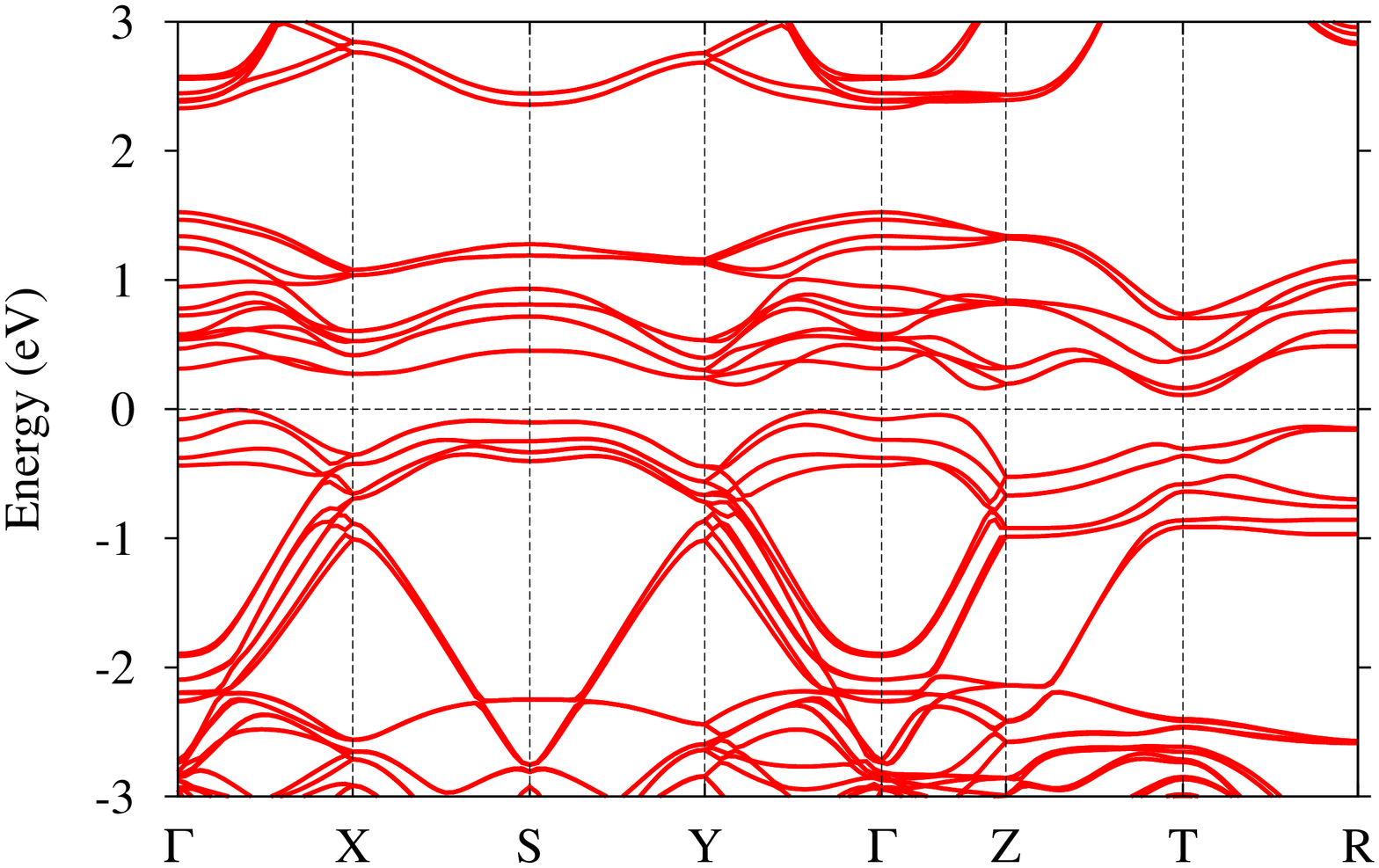,angle=-90,width=82mm}  
\psfig{figure=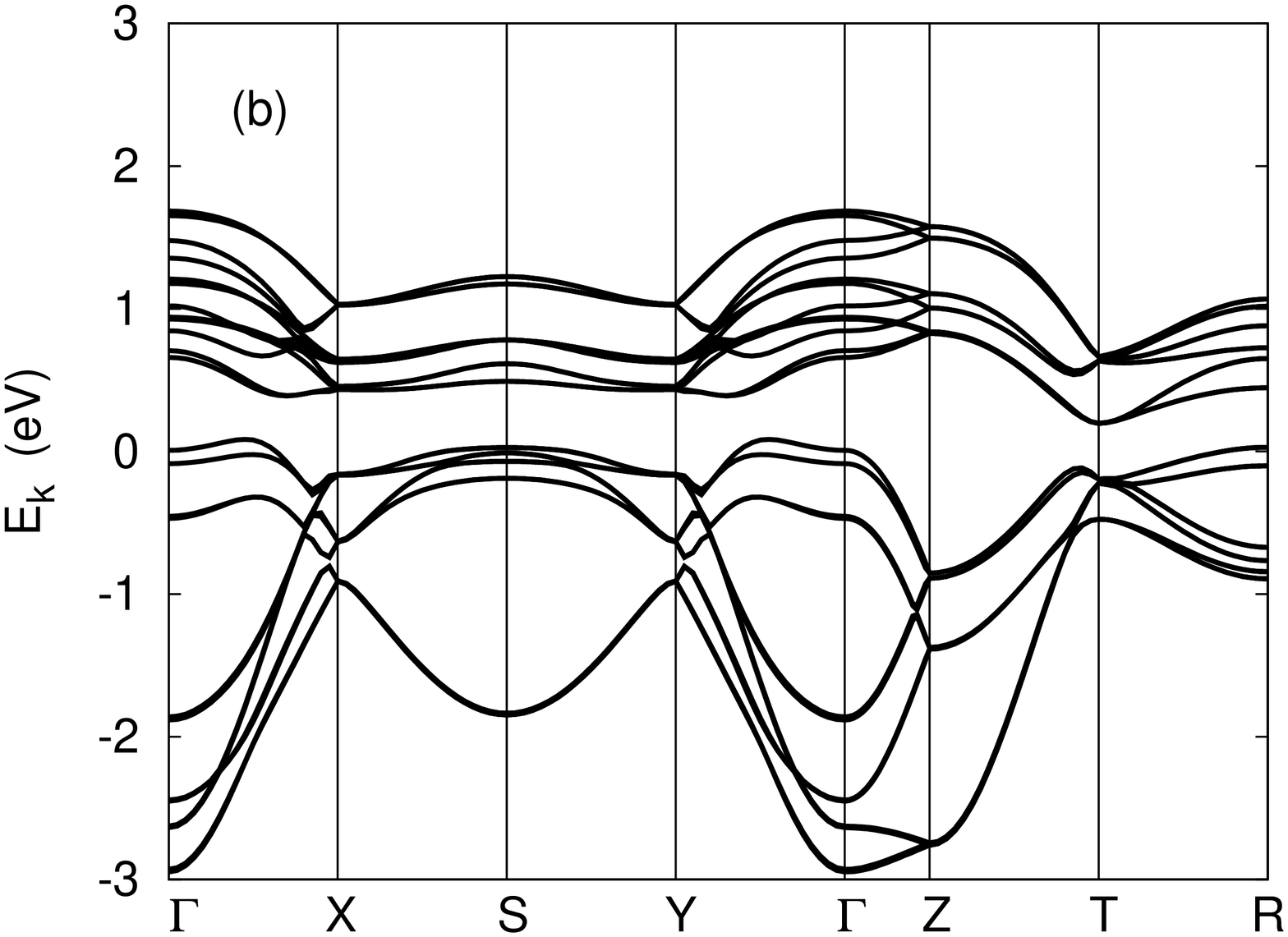,angle=-90,width=80mm}  
\caption{Electronic band structures in the AFM state of $\rm Na Os O_3$ obtained from (a) DFT calculation and (b) the three-orbital model, both with SOC and structural distortion.} 
\label{band_distort}
\end{figure}

Figure \ref{band_distort} shows the electronic band structures for the distorted case from DFT calculation and from the three-orbital model. The momenta $k_x,k_y,k_z$ are in units of $a^{-1}$, $c^{-1}$, and $b^{-1}$, respectively, for the distorted structure (DFT calculation). As seen from the band structure plots [Figs. \ref{band_undis}(b) and \ref{band_distort}(a)], we indeed find that octahedral rotations have significant effects on the DFT band dispersion and band width for the distorted structure. The bands near the Fermi energy become narrower and flatter in the distorted structure resulting in an insulating gap of $\sim 0.1$ eV. The positions of the conduction band (CB) minimum and the valence band (VB) maximum also shift. For example, there are multiple VB maxima away from the $\Gamma$ point where the maximum lies for the undistorted structure.

In order to reproduce essential features and the overall energy scale of the DFT band structure in our three-orbital model, we have taken hopping parameter values as listed in Table \ref{table-2}, staggered field $\Delta = 0.8$, and $\lambda=0.9$, with energy scale $|t_{1(s)} ^{(xy)}| = 420$ meV ($\Delta=0.34$ eV and $\lambda = 0.38$ eV). The structural distortion effect has been incorporated by introducing both orbital and directional asymmetries in the hopping parameters, as discussed above. The orbital mixing hopping terms generated due to octahedral rotations are discussed in the Appendix. It is interesting to note that for the same $\Delta$ value as for the undistorted (metallic) case [Fig. \ref{band_undis}(d)], the distortion induced band flattening near $E_{\rm F}$ results in an insulating state [Fig. \ref{band_distort}(b)].     

\begin{table}[b]
\centering
\caption{Hopping parameter values in the three-orbital-model for the distorted structure. The energy scale unit is $|t_{1(s)} ^{(xy)}|$ and $\alpha = xz, yz$.}
\label{table-2}
\begin{ruledtabular}
\begin{tabular}{l c  c  c  c  c  c  c  c  c  c  c  c  c  c}
$t_{1(s)} ^{(xy)}$ & $t_{1(s)} ^{(\alpha)}$ & $t_{1(d)} ^{(\alpha)}$ & $t_{{1\delta}(d)}^{(xy)}$ & $t_{{1\delta}(s)}^{(\alpha)}$ & $t_{2(s)} ^{(xy)}$ & $ t_{2(d)}^{(\alpha)}$ & $t_{3(s)} ^{(xy)}$ & $t_{m1}$ & $t_{m2}$ & $\epsilon_{xy}$ \\ 
\hline 
 -1.0 & -0.9 & -0.7 & 0.0 & -0.15 & 0.3 & 0.2 & 0.12 & 0.15 & 0.15 & 0.0 \\
\end{tabular} 
\end{ruledtabular}
\end{table}

The DFT bands clearly show that the VB peak at $\Gamma$ (undistorted structure) shifts to finite momentum (distorted structure). This is accompanied with significant band narrowing near the top of the VB [Figure \ref{band_distort}]. We show here that these features can be reproduced by introducing first neighbor hopping asymmetry in the $z$ direction. Consider the band dispersion term for the $xz$ and $yz$ orbitals: $\varepsilon_{\bf k}^{(\alpha)} = -2[t_{1(s)}^{(\alpha)} \cos(k_x/2) + t_{1(d)}^{(\alpha)} \cos{k_z}]$ along the $\Gamma \rightarrow X$ direction, where $k_z = \pi$ for the upper branch of the VB near the $\Gamma$ point. Since this term contributes to the AFM state energy as: $-\sqrt{(\Delta^2 + \varepsilon_{\bf k}^{(\alpha)2})}$, the VB peak near $\Gamma$ corresponds to $\varepsilon_{\bf k}^{(\alpha)}$ = 0. Now, for $t_{1(s)}^{(\alpha)} = t_{1(d)}^{(\alpha)}$ (cubic symmetry), the peak occurs at $k_x = 0$ ($\Gamma$ point), as seen for the undistorted structure [Fig. \ref{band_undis}]. However, 
$t_{1(d)}^{(\alpha)} < t_{1(s)}^{(\alpha)}$ for the peak to occur at finite $k_x $, as seen for the distorted structure [Fig. \ref{band_distort}]. So the band structure for the distorted case clearly indicates a first neighbor hopping asymmetry for the $xz$ and $yz$ orbitals, with a reduced out-of-plane hopping.

\begin{figure}
\vspace*{0mm}
\hspace*{0mm}
\psfig{figure=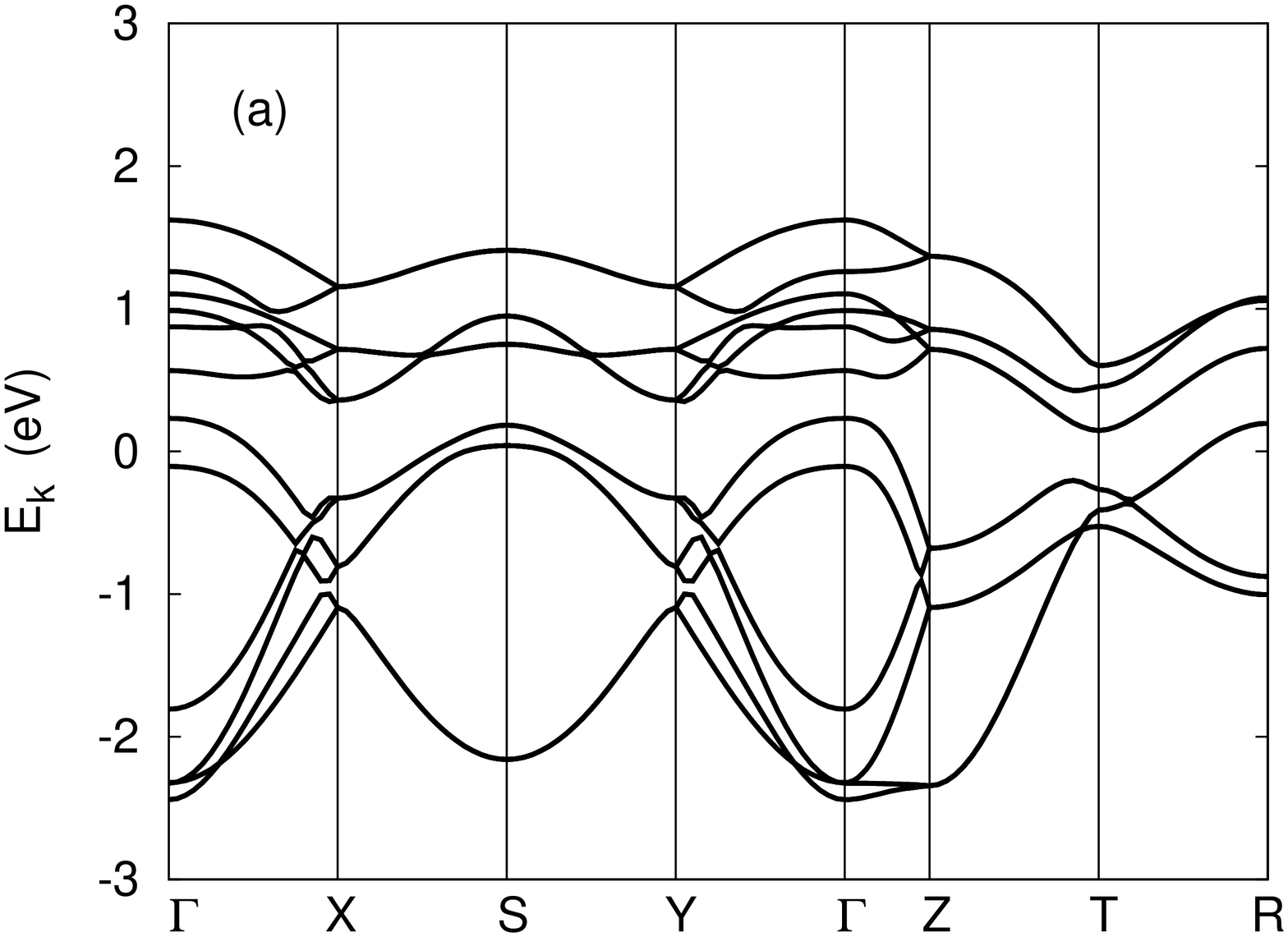,angle=-90,width=50mm}
\psfig{figure=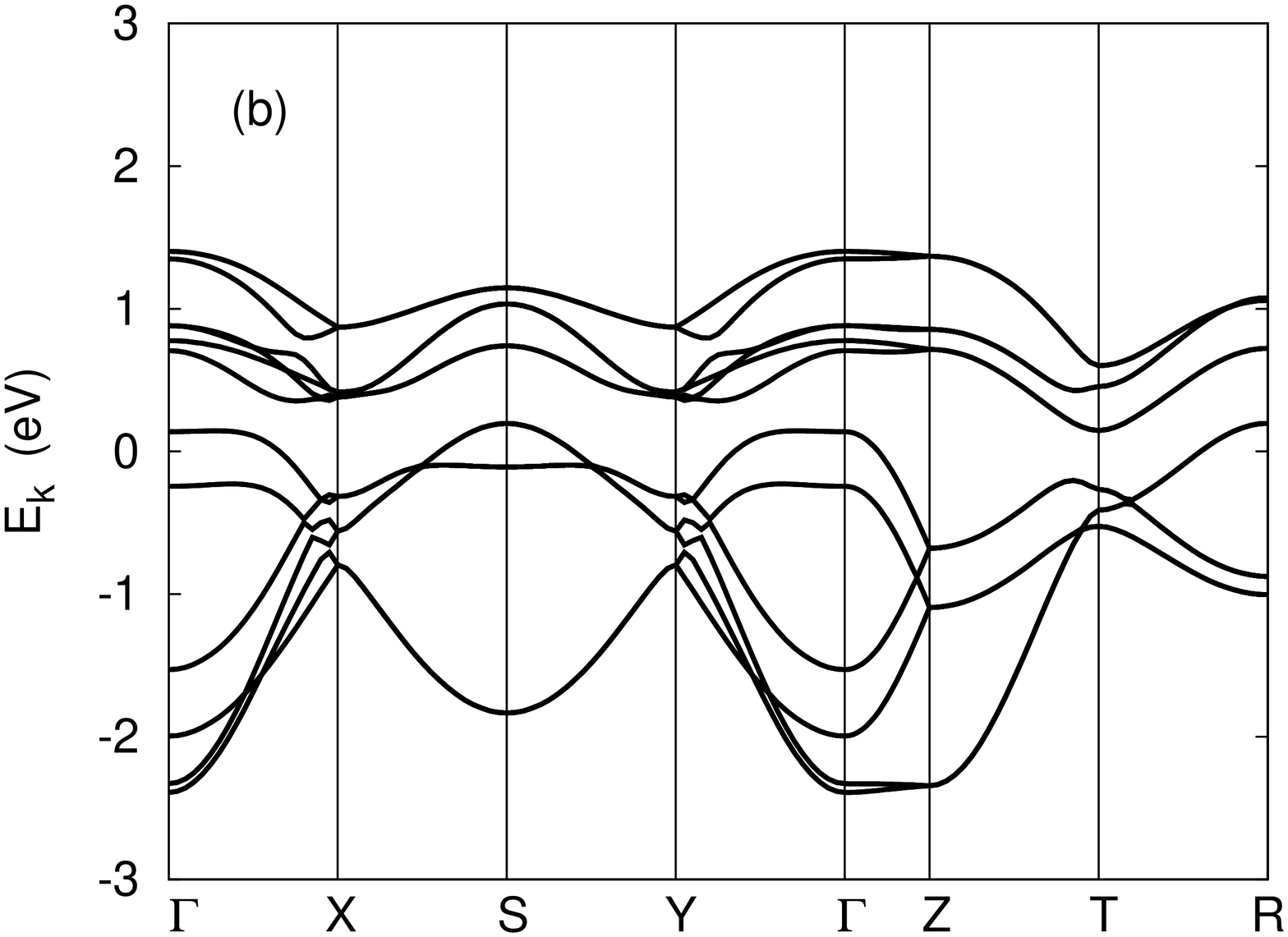,angle=-90,width=50mm} \\
\psfig{figure=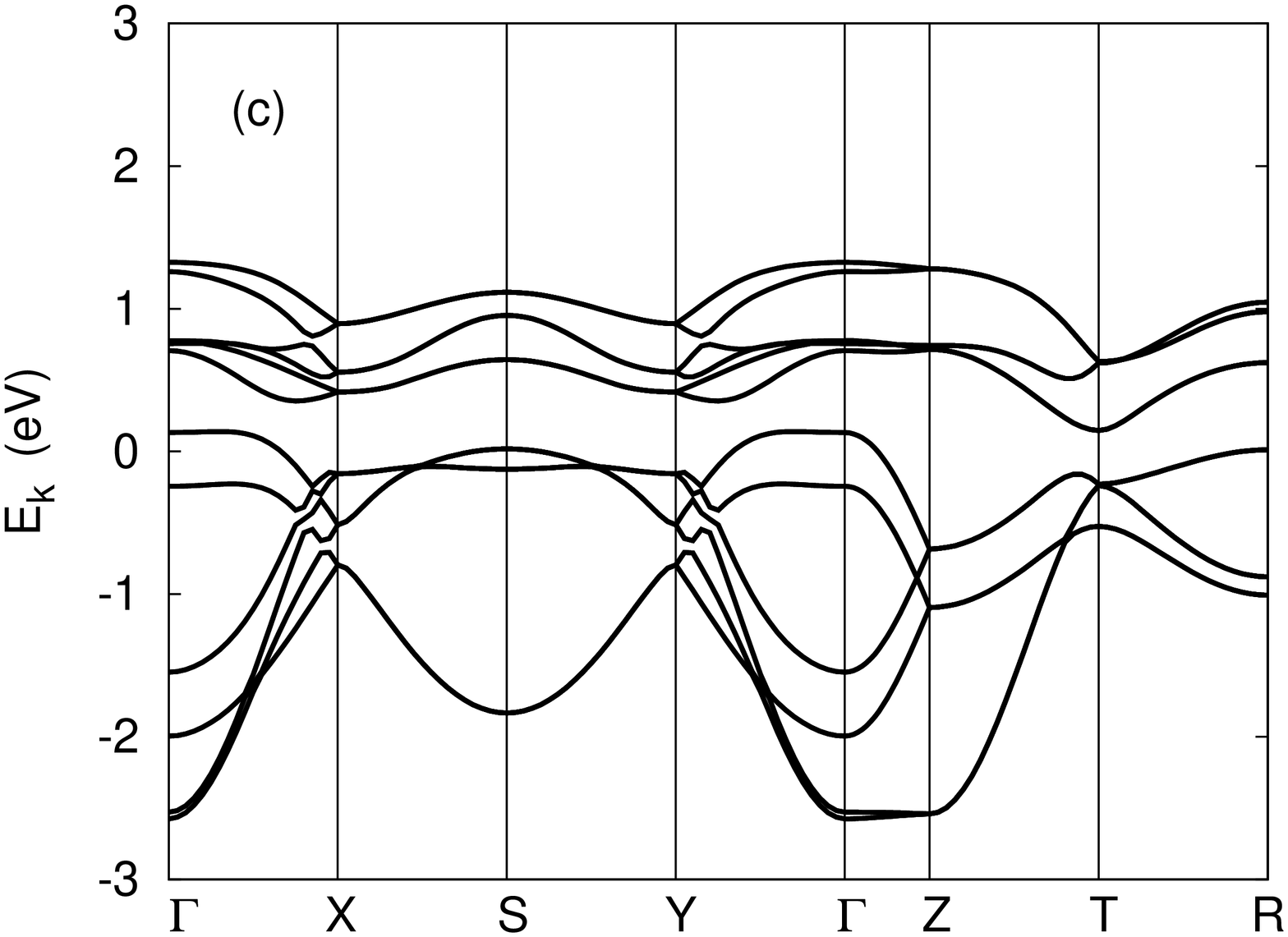,angle=-90,width=50mm}
\psfig{figure=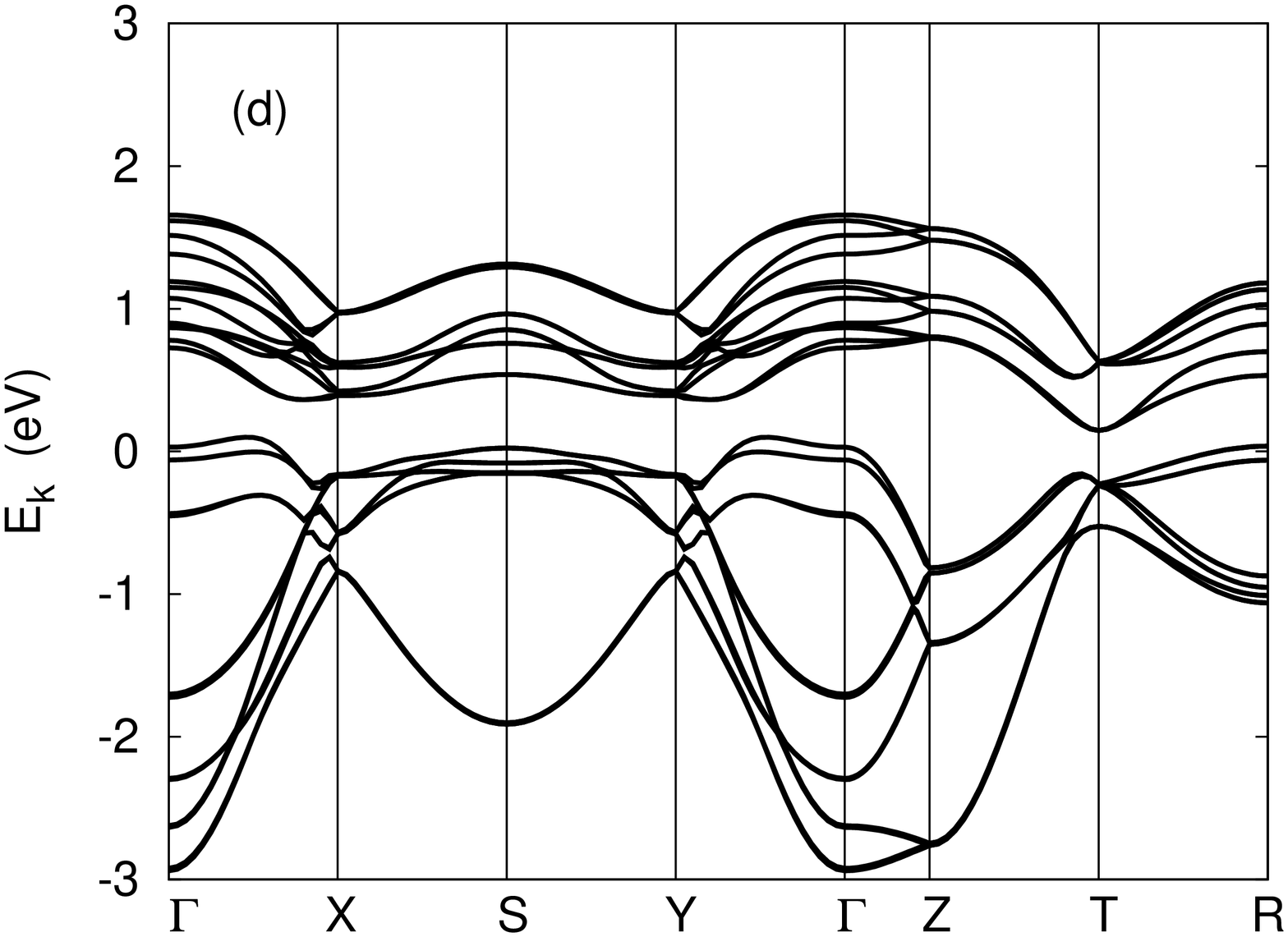,angle=-90,width=50mm}    
\caption{Evolution of the electronic band structure with the distortion-induced changes in the hopping terms. Starting from the undistorted case with SOC, the panels progressively correspond to (a) reduced $t_2$ for $xz/yz$, (b) reduced $t_{1(d)}$ for $xz/yz$, (c) including $t_{3(s)}^{(xy)}$ for $xy$, and (d) including both the mixing terms $t_{m1},t_{m2}$.} 
\label{evolution}
\end{figure}

The essential features distinguishing the distorted structure are: (i) overall energy scale reduction (from 500 meV to 420 meV), (ii) negative curvature near $\Gamma$ (resulting from $t_{1(d)}^{(\alpha)} < t_{1(s)}^{(\alpha)}$, $t_{2 (d)}^{(\alpha)} <  t_{2(s)}^{(xy)}$, orbital mixing), (iii) narrowing of $xz/yz$ bands near top of the VB (resulting from reduced $xz/yz$ hopping terms), and (iv) fine band splittings (due to the orbital mixing term $t_{m2}$). The series of panels in Fig. \ref{evolution} show the evolution of the band structure with the important changes in the hopping terms induced by the distortion. Starting from the undistorted case with SOC, the panels show (a) reduced positive curvature, (b) negative curvature near $\Gamma$ and band flattening, (c) energy lowering at S and R (all at top of VB), and (d) fine band splittings and enhanced negative curvature due to the orbital mixing terms.Among the broad features, $t_2$ (connecting same magnetic sublattice) controls the bandwidth asymmetry between the valence and conduction bands.

Some other fine features can be further improved by incorporating additional terms in the three band model. For example, the energy of the lower branch at $S$ can be pulled down by including negative 2$^{\rm nd}$ neighbor hopping $\tilde{t}_2$ ($\pi-\delta$ overlap) and positive 3$^{\rm rd}$ neighbor hopping $t_{3}$ ($\pi$ overlap). The $S-Y$ versus $S-X$ asymmetry (conduction band) as seen in DFT calculation is obtained by including hopping asymmetry between the $a$ and $c$ directions.

The dominant hopping paths between neighboring Os $d$ orbitals are provided by the intervening oxygen $p$ orbitals. The effects of structural distortion on the oxygen orbitals has been neglected in the transformation analysis of the hopping Hamiltonian matrix in the rotated basis (Appendix A). As both rotation and tilting also displace the oxygen ions, some of the $p \, d \, \pi$ type orbital overlaps are significantly modified. For example, for staggered octahedral rotation about the $z$ direction, the $d_{xy}$ orbital overlap between neighboring Os in the $x$ direction through the intervening O $p_y$ orbital is reduced due to the displacement of O along the $y$ direction.\cite{fang_PRB_2001} From our band structure comparison (Figs. \ref{band_undis} and \ref{band_distort}), the distortion-induced reduction in the hopping energy scale (from 500 to 420 meV) in our three-orbital model is a consequence of this structural distortion effect on the $p \, d \, \pi$ type orbital overlaps.

As seen from Fig. \ref{band_distort}, the distortion-induced bandwidth narrowing results in a marginally insulating system. Further reduction in $\Delta$ due to temperature will lead to formation of small electron (hole) pockets at the T ($\Gamma$) points, resulting in a continuous MI transition and magnetic transition due to the low DOS at the Fermi energy. This spin driven Lifshitz transition scenario has been recently proposed for the $\rm Na Os O_3$ compound, where the continuous metal-insulator transition is suggested to be associated with a progressive change in the Fermi surface topology.\cite{kim_PRB_2016} 

\section{Sublattice magnetization}
\begin{figure}
\vspace*{0mm}
\hspace*{0mm}
\psfig{figure=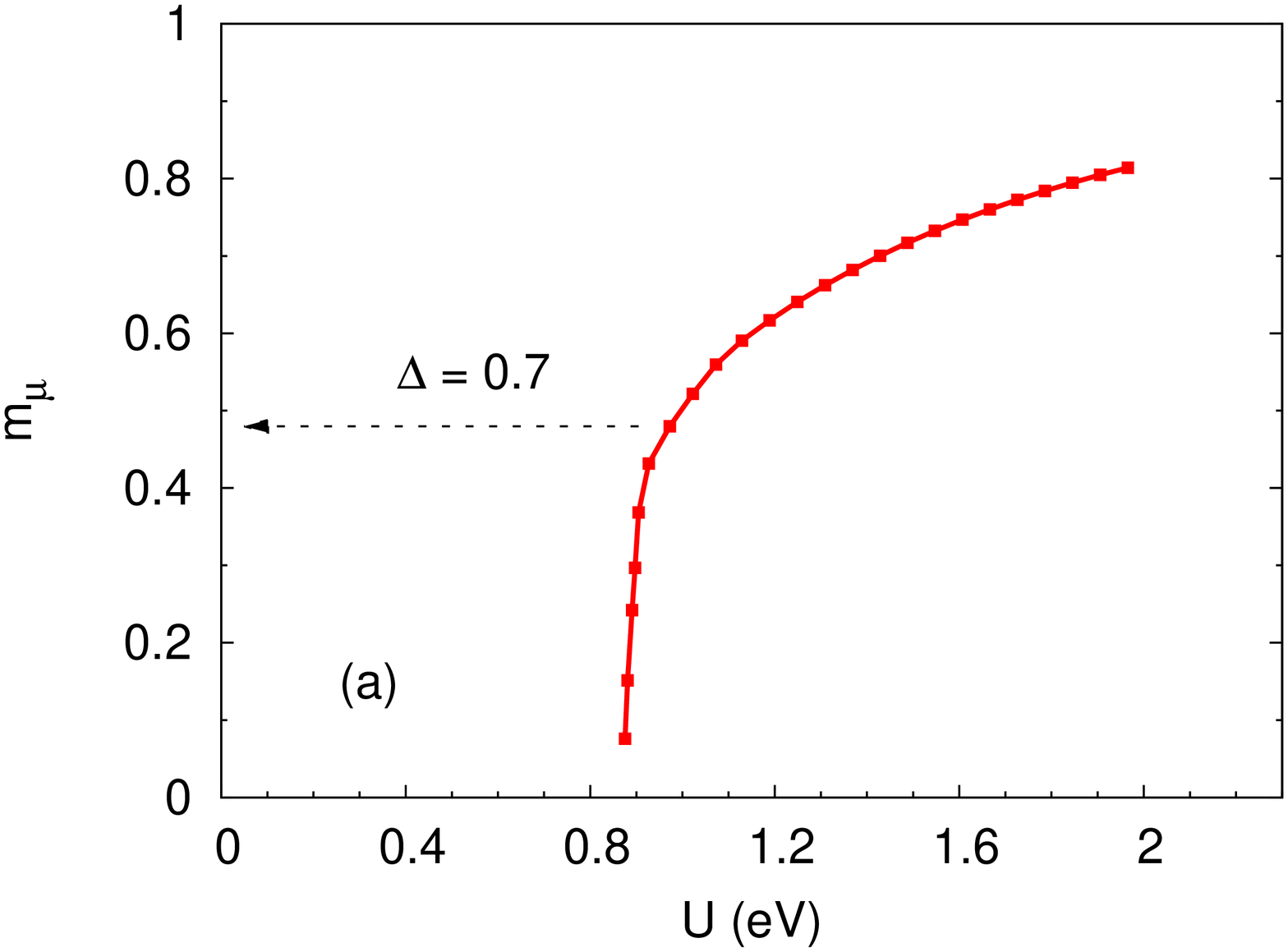,angle=-90,width=80mm} 
\psfig{figure=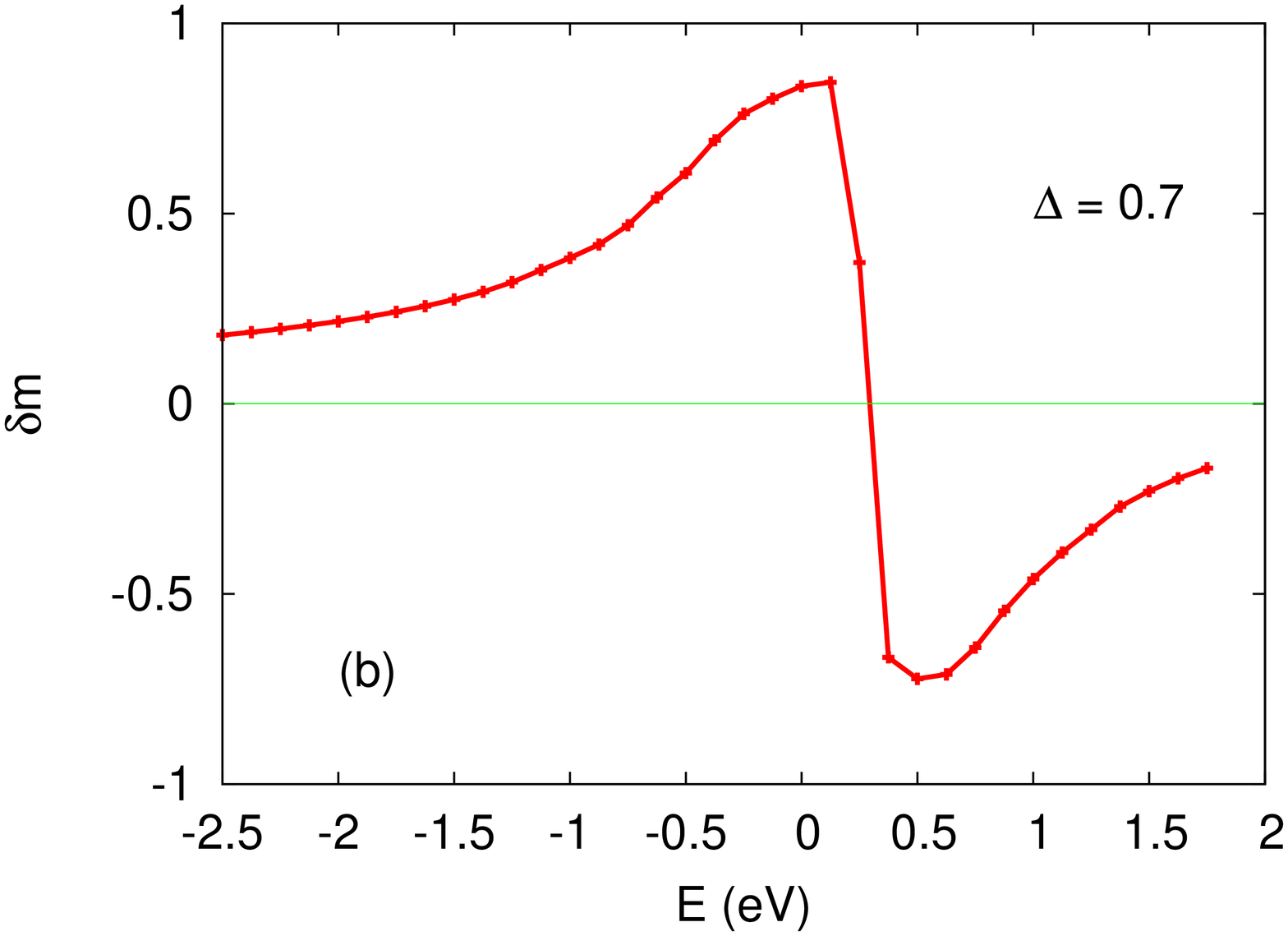,angle=-90,width=80mm} 
\caption{Variation of (a) staggered magnetization with $U$ and (b) energy resolved staggered magnetisation with energy in the three-orbital model. SOC and hopping parameter values correspond to Fig. \ref{band_undis} (d) for the undistorted case.} 
\label{mag_undis}
\end{figure}

In order to determine the $U$ values corresponding to Figs. \ref{band_undis} and \ref{band_distort} for the band structure within the three-orbital model, we have also evaluated the staggered magnetization in the AFM state from Eq. \ref{magneqn}. The ${\bf k}$ sum over the three-dimensional Brillouin zone was performed using a $20 \times 20 \times 20$ mesh. For the undistorted case, Fig. \ref{mag_undis} shows a strong variation of $m_\mu$ in the physically relevant $U$ range near $\Delta/|t_1| = 0.7$ as in Fig. \ref{band_undis}(d), highlighting the weak correlation in $\rm NaOsO_3$. The orbital moment $m_{\rm orb}$ (listed in Table III) was calculated using Eq. \ref{mag_orb} for the same interaction strength. 

\begin{table}[b]
\caption{Comparison of the magnetic moment values (in $\mu_{\rm B}$) calculated within DFT [LSDA + SO ($+U$)] and the three-orbital model.}
\begin{ruledtabular}
\begin{tabular}{lcccc}
Method & LSDA & +SO & +SO+U & Exp.\cite{calder_PRL_2012} \\ 
\hline
DFT (spin) &  0.66 & 0.22 & 0.96 & 1.01 \\
DFT (orbital) &  - & -0.03 & -0.08 & \\
Three-orbital model (spin / orbital) & - & - & 1.4 / 0.02 & \\
%Three-orbital model (orbital) & - & - & 0.04 &  \\
Atomic limit (spin / orbital) & - & - & 2.5 / 0.2 &  \\
%Atomic limit (orbital) & - & - & 0.2 &  \\
\end{tabular}
\end{ruledtabular}
\label{tab1}
\end{table}

As indicated by the arrow in Fig. \ref{mag_undis}, we obtain $m_\mu \approx 0.48$ (for all three orbitals), yielding $m_{\rm spin} = \sum_\mu m_\mu \approx 1.4$. This is somewhat larger than our DFT result (Sec. II) and the experimental value of about 1 $\mu_{\rm B}$ due to neglect of the strong $\rm Os-O$ hybridization which reduces the electron density on Os in the $t_{2g}$ sector. From Eq. \ref{selfcon}, and assuming $J_{\rm H} = U/4$, the self-consistently determined value $U \approx 1$ eV (without Hund's coupling, $U \approx 1.5$ eV) is within the estimated range for osmates.\cite{jung_PRB_2013,shi_PRB_2009} 
For the distorted structure (Fig. \ref{band_distort}), we obtain $m_{xz} = m_{yz} \approx 0.53$ and $m_{xy} \approx 0.47$ for the staggered magnetizations and again $U \approx 1.5$ eV (without $J_{\rm H}$). This small orbital disparity is consistent with our finding of an effective bandwidth narrowing for the $xz,yz$ orbitals relative to the $xy$ orbital from our electronic band structure investigation in Sec. III. 

Comparison of the magnetic moment values as obtained from the different methods used in our work is shown in Table \ref{tab1}. The spin and orbital moments in the atomic limit are obtained from Fig. \ref{atomic} (b) with $2\Delta \approx U$ and $\Delta / \lambda \sim 1$, corresponding to the physically relevant parameters. The orbital moment $m_{\rm orb}$ can be analytically shown to be identical to $n_\downarrow$ (Fig. \ref{atomic}). The spin and orbital moments exhibit opposite behavior with interaction strength, reflecting progressive loss of spin-orbital entanglement. The large moment reduction in the band limit compared to the atomic limit is due to the highly itinerant character of the system. 

Variation of the energy resolved staggered magnetization $\delta m$ defined below Eq. \ref{magneqn} (with $\delta E_{\rm band} = 0.15$ eV) is shown in Fig. \ref{mag_undis}(b). The strongly magnetic character of states near the Fermi energy is characteristic of the weakly correlated AFM state. Significant reduction in $m$ at finite temperature due to thermal electronic excitation across the Fermi energy should be the dominant (Slater type) demagnetization mechanism in this compound having a large magneto-crystalline anisotropy gap. 

\section{Magnetic anisotropy energy}
In this section we will study the SOC-induced magnetic anisotropy within the three-orbital model corresponding to the distorted structure. In this context, we note here that the $xy$ orbital density ($n_{xy}$) exhibits an intrinsic SOC-induced reduction on rotating the staggered field orientation from the $z$ direction ($\theta = 0$) to $x/y$ direction ($\theta = \pi/2$), accompanied with a corresponding increase in the $yz,xz$ orbital densities. This suggests that a positive energy offset $\epsilon_{xy}$ (or, equivalently, negative energy offset for the $yz,xz$ orbitals, or a combination of both) should  contribute to the magnetic anisotropy energy (MAE), resulting in easy $x-y$ plane anisotropy. We have therefore included a small positive $\epsilon_{xy}$ (possibly arising from tetragonal distortion of the $\rm OsO_6$ octahedra) which couples with the SOC-induced reduction in $n_{xy}$. Another microscopic factor which has been found to contribute to the SOC-induced MAE within a simplified three-orbital model is a relative bandwidth narrowing of the $xz/yz$ bands compared to the $xy$ band,\cite{osmate_new} which is indeed confirmed from the detailed band structure comparison for the distorted structure (Sec. III). 

\begin{figure}
\vspace*{0mm}
\hspace*{0mm}
\psfig{figure=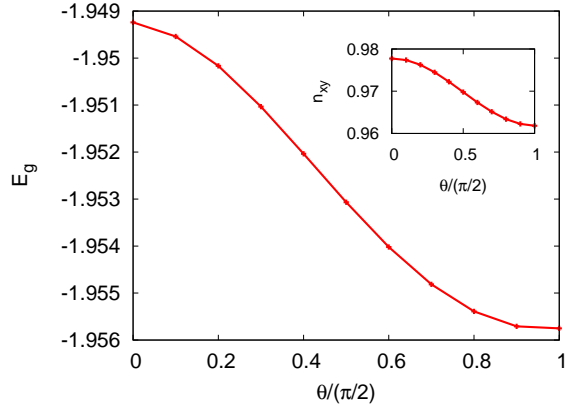,angle=0,width=80mm}  
\caption{SOC-induced magnetic anisotropy, as shown by the dependence of the electronic ground state energy (per state) with the staggered field orientation $\theta$  from the $z$ axis. Inset shows the small reduction in the $xy$ orbital density with orientation $\theta$.} 
\label{mag_aniso}
\end{figure}

Fig. \ref{mag_aniso} shows the ground state energy variation as the staggered field orientation is rotated from $z$ direction $(\theta = 0)$ to $x/y$ direction $(\theta = \pi/2)$. The hopping parameters considered (Table II) are same as for the electronic band structure (Fig. \ref{band_distort}). A slightly larger SOC value $\lambda = 1.1$ was taken to enhance the magnetic anisotropy effect, with the staggered field magnitudes $\Delta_{yz} = \Delta_{xz} = 1.1$ and $\Delta_{xy} = 0.93$ to preserve the AFM insulating state in the entire $\theta$ range, which yields $U_\mu \approx 3.5$ for all three orbitals ($J_{\rm H} = 0$). The calculated MAE value $\Delta E_g = E_g (z) - E_g (x) \approx 0.007$ per state with $\epsilon_{xy}=0.3$ ($\approx 0.005$ without $\epsilon_{xy}$). Using the overall energy scale $t_1 = 420$ meV, we obtain the effective single-ion anisotropy energy:
\begin{equation} 
\Delta E_{\rm sia} = 3 \times 0.007 \times 420\; {\rm meV} \approx 9 \; {\rm meV} 
\end{equation}
where the factor 3 accounts for the conversion from average energy per state to average energy per ion (corresponding to the three $t_{2g}$ orbitals per Os). The above value is in agreement with that obtained from the single-ion anisotropy term $\Delta E_{\rm sia} = DS_{iz} ^2$ = 9 meV for $D = 4$ meV and $S = 3/2$, as considered phenomenologically in localized spin models.\cite{calder_PRB_2017}

Both microscopic factors which contribute to the SOC-induced MAE as discussed above are dependent on the Coulomb interaction $U$. Due to progressive suppression of the SOC-induced spin-orbital entanglement with increasing $U$ (Fig. \ref{atomic}), the densities for all three orbitals will asymptotically approach 1 in the strong coupling limit ($U/t \rightarrow \infty$), and become independent of the staggered field orientation. Similarly, the effect of the relative bandwidth narrowing of the xz/yz bands compared to the xy band, which produces a small but crucial magnetic moment difference in the weak correlation limit,\cite{osmate_new} will be suppressed as the magnetic moments for all three orbitals will saturate to 1. Thus, the weak correlation term plays a key role in the generation of large magnetic anisotropy energy.

%A detailed study of the SOC-induced magnetic anisotropy has been recently carried out within a simplified three-orbital model at half filling.\cite{osmate_new} Two essential factors were identified which contribute to the magnetic anisotropy: i) hopping asymmetry with reduced first-neighbor hopping for the $yz,xz$ orbitals compared to the $xy$ orbital and ii) positive energy offset $\epsilon_{xy}$ for the $xy$ orbital. With these two factors included, the calculated spin wave energy gap obtained ($ \sim 50$ meV)  is in close agreement with the experimental measurements. The small hopping asymmetry ($t_1 ^{yz,xz} < t_1 ^{xy}$) results in SOC-induced anisotropic magnetic interactions. On the other hand, positive energy offset $\epsilon_{xy}$ couples with the intrinsic SOC-induced reduction in the $xy$ orbital density on rotating the orientation of the staggered field from $z$ to $x/y$ direction, resulting in easy $x-y$ plane anisotropy. Our detailed electronic band structure comparison with the DFT result (Sec. III) clearly shows the hopping asymmetry feature. Including a positive energy offset $\epsilon_{xy}$ should further enhance the magnetic anisotropy energy.  
 
%magnetic moment asymmetry ($m_{yz,xz} > m_{xy}$) and 

%We have therefore included both these factors in the three-orbital model: the small first-neighbor hopping asymmetry ($t_1 ^{yz,xz} < t_1 ^{xy}$) and positive energy offset for the $xy$ orbital with respect to the degenerate $xz/yz$ orbitals (possibly corresponding to the tetragonal distortion of the $\rm OsO_6$ octahedra).

\section{Conclusions}

The effects of SOC and octahedral rotations on the electronic band structure of $\rm NaOsO_3$ were investigated using density-functional methods and a minimal three-orbital model. Our DFT results show that $\rm NaOsO_3$ is an AFM band insulator with a small gap ($\sim$ 0.1 eV) consistent with experiments and previous calculations. Octahedral rotations have a significant effect on the band structure. While a negative indirect band gap is obtained for the undistorted structure, a small band gap opens up for the distorted structure with the octahedral rotations, along with significant bandwidth reduction and flattening of certain bands near the Fermi energy. The calculated magnetic moment per Os atom is in good agreement with the experimental value.

Complementary to the DFT-based approaches, the tight-binding approach has provided valuable insight into the electronic and magnetic properties of the system, as summarized below. Essential features of the DFT band structure for both the undistorted and distorted structures were reproduced within the minimal three-orbital model. Among the broad features, the bandwidth asymmetry between the valence and conduction bands arises typically in the AFM state from the hopping terms ($t_2, t_3$) connecting the same magnetic sublattice. Orbital and directional asymmetry in the hopping terms associated with the cubic symmetry breaking resulting from the structural distortion was clearly indicated. The band structure comparison further showed that some minute and robust features in the DFT band structure such as the fine band splitting and the conduction band features in the X-S-Y momentum range can be ascribed to the orbital mixing terms arising from the octahedral rotation and tilting, as obtained from the Hamiltonian matrix transformation. The orbital mixing terms also contribute significantly to the negative band curvature near $\Gamma$. 

%The orbitally asymmetric bandwidth reduction is important for the SOC-induced magneto-crystalline anisotropy in this compound.

The behaviour of staggered magnetization with interaction strength in the three-orbital model and the strongly magnetic character of states near the Fermi energy were found to be characteristic of weakly correlated AFM state. Nearly 50\% additional contribution to the staggered field is provided by the Hund's coupling term, highlighting its importance in stabilizing the barely insulating AFM state in $\rm NaOsO_3$, along with the bandwidth narrowing due to structural distortion. The calculated magnetic anisotropy energy was found to be in agreement with the single-ion anisotropy term as considered phenomenologically in localized spin models.

\appendix*
\makeatletter		
\renewcommand*\env@matrix[1][*\c@MaxMatrixCols c]{%
\hskip -\arraycolsep
\let\@ifnextchar\new@ifnextchar
\array{#1}}
\makeatother

\section{Transformation of the d electron tight-binding hopping matrix elements for osmates under rotation and tilting}

\maketitle
 
\begin{figure} [ht]
\includegraphics[scale=0.4] {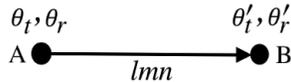} 
\caption {Two neighboring Os sites, with the angles describing the octahedral rotation with respect to the crystalline axes. 
Direction cosines are denoted by $l$, $m$, and $n$.}
\label {fig-AB}
\end{figure}

In this Appendix, we provide the expressions for the tight-binding (TB)  hopping integrals between the d orbitals located on two Os atoms. The atoms denoted by A and B are  separated by a distance vector with direction cosines ($l$, $m$,  and $n$) (Fig. \ref{fig-AB}), and the local coordinate axes (with respect to which the orbital lobes are defined) are rotated with respect to the crystalline axes. The local axes  point towards the O atoms on the OsO$_6$ octahedra. 

\begin{figure} [t]
\includegraphics[scale=0.35] {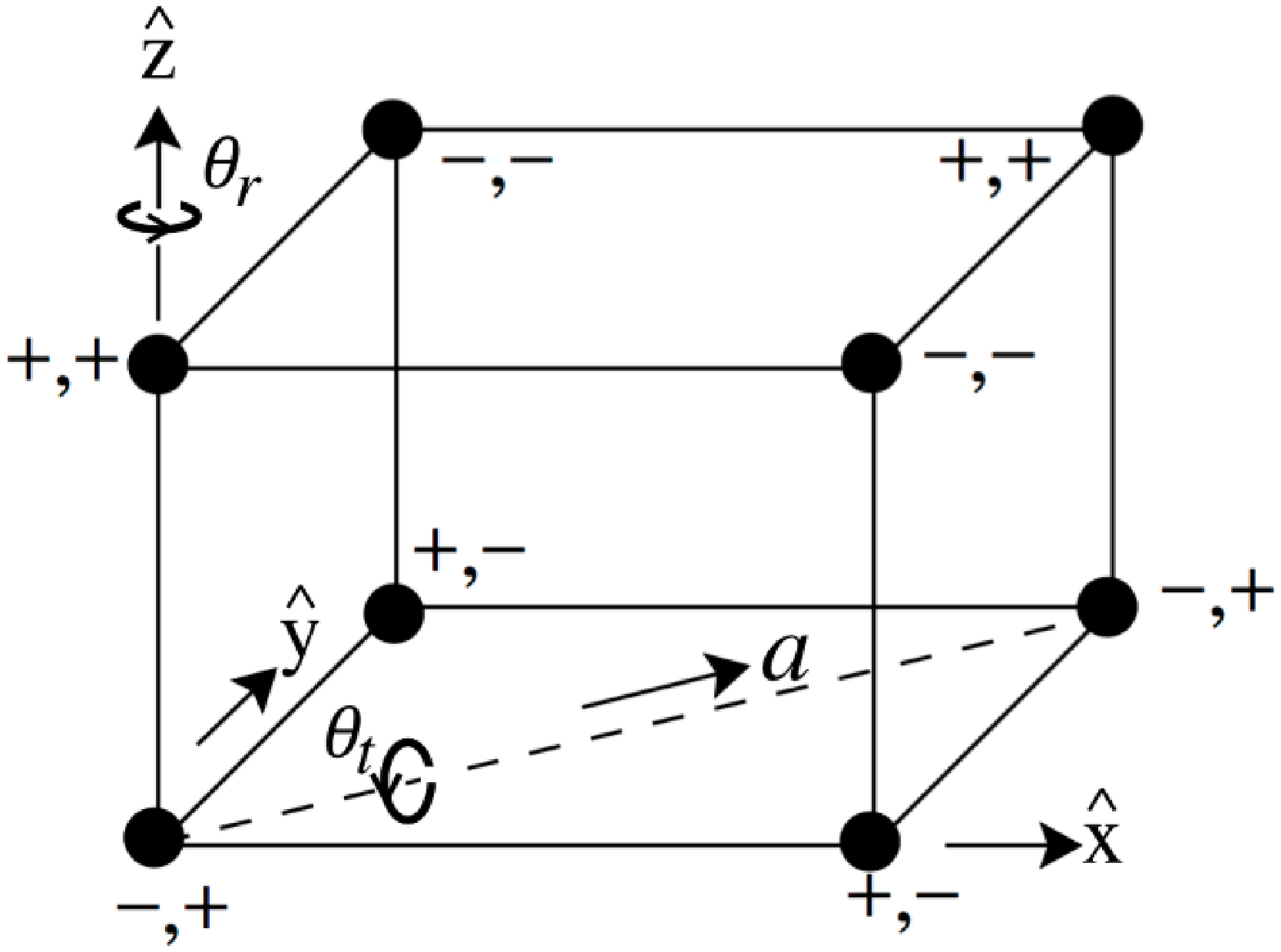} 
\includegraphics[scale=0.15] {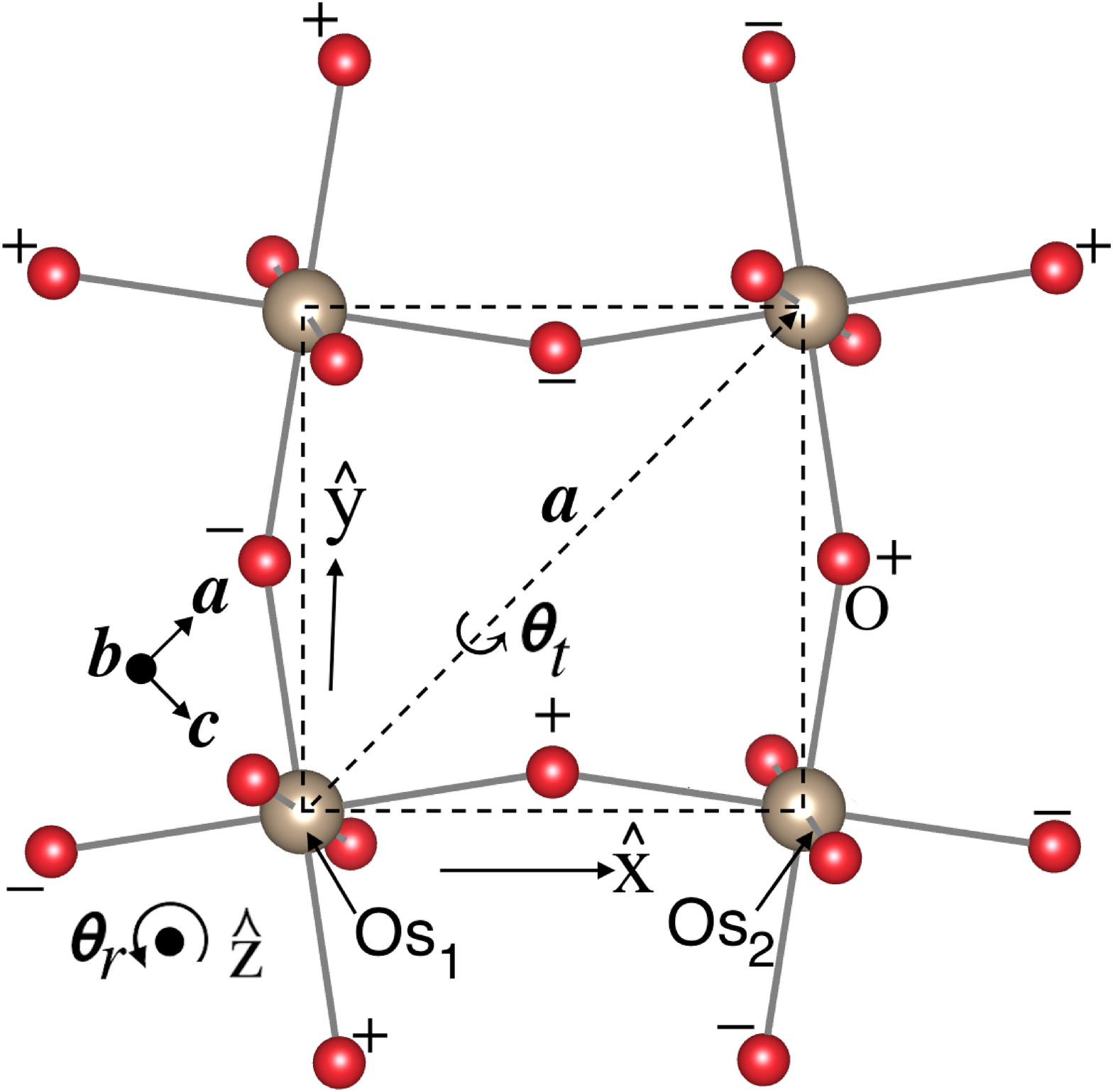} 
\caption{Tilt and rotation of the octahedra at the Os sites in NaOsO$_3$ ({\it left}). The two signs on each site indicate the sense of the  tilt and the rotation, counter clockwise ($+$) or clockwise ($-$). The full rotation is described by  the tilt $\theta_t$ (rotation about the $a$ axis), followed by the rotation $\theta_r$ (with respect to octahedral $\hat z$ axis). Octahedral orientations on the $ac$ plane ({\it right}); signs indicate the location of oxygen atoms above ($+$) or below ($-$) the $xy$ plane.}
\label{fig-cube}
\end{figure}

In the osmates, the octahedra are more or less undistorted, but they are rotated with respect to the crystalline axes. The rotation can be described as a tilt and a rotation, which vary from site to site  as indicated in Fig. (\ref{fig-cube}). The local octahedral axes are obtained by first rotating the crystalline axes counterclockwise by the angle $\theta_t$ about the diagonal direction indicated by $a$ (this is referred to as tilting), followed by a rotation of $\theta_r$ about the $\hat z$ axis. These axes are fixed for every Os atom as are the magnitudes of the angles $\theta_t$ and $\theta_r$ (both about 11$^\circ$ for NaOsO$_3$), but the rotations and tilts are either counterclockwise ($+$) or clockwise ($-$) as indicated by the pair of signs next to each Os atom in Fig. (\ref{fig-cube}). For example the `$- +$' next to the Os atom at the origin means that the tilt and rotation angles there are given by ($-\theta_t$ and $+\theta_r$).

We follow the standard convention that the function contours are rotated and the coordinate axes (or the basis vectors) always stay fixed (active rotation). 

Any rotation can be expressed in terms of the Euler angles, $\alpha$, $\beta$, and $\gamma$, for which we follow Rose's definition.\cite{Tinkham} The sequence of the three rotations are: first rotate by $\gamma$ about the $z$ axis, then by $\beta$ about $y$, and finally by $\alpha$ about $z$ again, all in the original fixed coordinate system. The same transformation may be obtained by rotating in the reverse order about the intermediate axes, viz., first $\alpha$ about the original $z$ axis, then by $\beta$ about the intermediate axis $y^\prime$, and finally by $\gamma$ about $z^{\prime \prime}$. 
Clearly, if $R_x (\theta)$ rotates the function by $\theta$ about $x$ axis, then the net effect of the  rotation  is given by the matrix $R(\alpha, \beta, \gamma) \equiv R_z (\alpha) R_y (\beta) R_z (\gamma)$ (or, equivalently by $R_{z^{\prime \prime} } (\gamma) R_{y^\prime} (\beta) R_z (\alpha)$, which is easier to visualize).

For the osmates, the final rotation matrix may also be generated by the sequence of the tilt and the rotation, viz., $R(\alpha, \beta, \gamma) = R_z (\theta_r) R_a (\theta_t)$. All angles vary from one Os atom to another, as indicated in Fig. (\ref{fig-cube}), and they are related by the expressions: $\alpha = -45^\circ +\theta_r$, $\beta = \theta_t$, and $\gamma = 45^\circ$.

Rotations don't mix functions with different angular momenta $L$, and therefore the d orbitals  ($ L = 2$) transform among one another, the transformation determined by a site-dependent $5 \times 5$ rotation matrix. We denote the unrotated  orbitals as $|\alpha \rangle$ ($ xy, yz, zx, x^2 - y^2,$ and $z^2 -1 $, in that order), rotated orbitals by $|\alpha^\prime \rangle$ on site A and $|\alpha^{\prime \prime} \rangle$ on site B, the corresponding rotation matrices by $R^\prime$ and $R^{\prime \prime}$, so that $|\alpha^\prime \rangle  = \hat R^\prime |\alpha \rangle$ and $|\alpha^{\prime \prime} \rangle  = \hat R^{\prime \prime}|\alpha \rangle$.
Then the hopping integrals between the rotated orbitals are given by
$\widetilde {H}_{\alpha^\prime \beta^{\prime \prime}} \equiv    \langle \alpha^{\prime } | H | \beta^{\prime \prime} \rangle = 
\langle \alpha | R^{\prime T} HR^{\prime \prime} |\beta \rangle$, or
\begin{equation}
\widetilde {H} =  R^{\prime T}  H R^{\prime \prime}.
\label{H-tilde}
\end{equation}

%: Tilt and Rotation matrices
{\it Rotation matrix for the Osmates} -- It is convenient to express the total rotation matrices for the d orbitals on each Os site in terms of the two individual rotations, so that
\begin{equation}
R (\theta_r, \theta_t) = R_z (\theta_r) R_a (\theta_t),
\label{RR}
\end{equation}
the two angles being positive or negative depending on the Os site. For pure rotations about $\hat z$, the Euler angles are simply $(\alpha, \beta, \gamma) = (\theta_r, 0, 0)$, while for pure tilt (rotation about $a$), they are  $(\alpha, \beta, \gamma) = (-45^\circ, \theta_t, 45^\circ)$. Using these angles and the expression for $R_2$ (Eq. \ref{R2}), we readily get the results:
\begin{widetext}
%:------ Rotation R_r
  \begin{eqnarray}
 R_z (\theta_r) =
      \bordermatrix{ & xy & yz &xz &x^2-y^2&3z^2-r^2 \cr
     xy~& \cos2\theta_r & 0 & 0 & \sin2\theta_r & 0\cr
     yz&  0 & \cos\theta_r & \sin\theta_r & 0 & 0\cr
     xz&  0 & -\sin\theta_r & \cos\theta_r& 0 & 0\cr
      x^2-y^2&-\sin2\theta_r & 0 & 0 & \cos2\theta_r & 0 \cr
     3z^2-r^2& 0 & 0 & 0 & 0 &1  }
 \end{eqnarray} %

\ \\

%------ Tilt matrix R_t
{\scriptsize
\begin{eqnarray}
 R_a (\theta_t) &=&
 \begin{pmatrix}
        \frac{1}{2}(1+\cos^2\theta_t) & \frac{1}{2\sqrt{2}}\sin2\theta_t & \frac{-1}{2\sqrt{2}}\sin2\theta_t & 0 &-\sqrt{\frac{3}{4}}\sin^2\theta_t  \\      
        - \frac{1}{2\sqrt{2}}\sin2\theta_t & \frac{1}{2}(2\cos^2\theta_t+\cos\theta_t-1)& \frac{-1}{2}(2\cos^2\theta_t-\cos\theta_t-1) & \frac{-1}{\sqrt{2}}\sin\theta_t & -\sqrt{\frac{3}{8}}\sin2\theta_t\\
        
        \frac{1}{2\sqrt{2}}\sin2\theta_t &\frac{-1}{2}(2\cos^2\theta_t-\cos\theta_t-1) &\frac{1}{2}(2\cos^2\theta_t+\cos\theta_t-1) &\frac{-1}{\sqrt{2}}\sin\theta_t &\sqrt{\frac{3}{8}}\sin2\theta_t    \\
        
        0 & \frac{1}{\sqrt{2}}\sin\theta_t & \frac{1}{\sqrt{2}}\sin\theta_t & \cos\theta_t & 0\\
        
        -\sqrt{\frac{3}{4}}\sin^2\theta_t &\sqrt{\frac{3}{8}}\sin2\theta_t  &-\sqrt{\frac{3}{8}}\sin2\theta_t & 0 & 1-\frac{3}{2}\sin^2\theta_t 
       \end{pmatrix}.
 \end{eqnarray}}
 \end{widetext}
 %------------------------------------------
So the combined effect of tilting and rotation is given by the product of these matrices, and keeping terms linear in the angles, we get
%
%----Total rotation matrix R -linear terms only-------------
\begin{eqnarray}
R  (\theta_r,  \theta_t) =      \begin{pmatrix}
1 &  \frac{1}{\sqrt{2}}  \theta_t&  -\frac{1}{\sqrt{2}}\theta_t &2\theta_r & 0 \\

- \frac{1}{\sqrt{2}}  \theta_t& 1    & \theta_r  & -\frac{1}{\sqrt{2}}\theta_t   & -\sqrt \frac{3}{2} \theta_t   \\

\frac{1}{\sqrt{2}}\theta_t    & -\theta_r   & 1  & -\frac{1}{\sqrt{2}}\theta_t & \sqrt \frac{3}{2} \theta_t\\

-2\theta_r &  \frac{1}{\sqrt{2}}\theta_t  &  \frac{1}{\sqrt{2}}\theta_t  & 1 &   0\\

 0 &\sqrt{\frac{3}{2}}\theta_t  &-\sqrt{\frac{3}{2}}\theta_t & 0 & 1 
\end{pmatrix}.
\label{mixing}
\end{eqnarray}

%: Rotated TB Hopping 

{\it TB hopping in the rotated basis} -- The TB matrix elements between the d orbitals on two different sites may be obtained by a straightforward use of Eqs. (\ref{H-tilde}) and (\ref{RR}). We illustrate this for the nearest-neighbor hopping from the Os at the origin to the Os located along the $+$ $\hat x$, $\hat y$, or $\hat z$ directions, for which the TB matrix elements in the unrotated basis can be obtained from Harrison's book.\cite{Harrison} These are
\begin{equation}
H_{100}=\begin{pmatrix}
V_{\pi} & 0 & 0 & 0 &0\\

0& V_{\delta} & 0 & 0 &0\\

0 & 0 & V_{\pi} & 0 &0\\

0 & 0 &0& \frac{1}{4}(3V_{\sigma}+V_{\delta}) & -\frac{\sqrt{3}}{4}(V_{\sigma}-V_{\delta})\\

0 & 0 & 0 &-\frac{\sqrt{3}}{4}(V_{\sigma}-V_{\delta}) & \frac{1}{4}(V_{\sigma}+3V_{\delta})
\end{pmatrix}, \nonumber
\end{equation}
\begin{equation}
H_{010}=\begin{pmatrix}
V_{\pi} & 0 & 0 & 0 &0\\

0& V_{\pi} & 0 & 0 &0\\

0 & 0 & V_{\delta}& 0 &0\\

0 & 0 &0 & \frac{1}{4}(3V_{\sigma}+V_{\delta}) & \frac{\sqrt{3}}{4}(V_{\sigma}-V_{\delta})\\

0 & 0 & 0 &\frac{\sqrt{3}}{4}(V_{\sigma}-V_{\delta}) & \frac{1}{4}(V_{\sigma}+3V_{\delta})
\end{pmatrix}, \nonumber
\end{equation} 
\vspace{1mm}
\begin{equation}
{\rm and} \hspace{5mm}
H_{001}=\begin{pmatrix}
V_{\delta} & 0 & 0 & 0 &0\\ 

0& V_{\pi} & 0 & 0 &0\\

0 & 0 & V_{\pi} & 0 &0\\

0 & 0 & 0 & V_{\delta} &0\\

0 & 0 & 0 &0 &V_{\sigma}
\end{pmatrix}.
\end{equation}
 %--- End  --- Harrison Hamiltonians  

In the following, we neglect $V_{\delta}$, which is much smaller compared to $V_{\sigma}$ and $V_{\pi}$, but can be included  without any problem. The TB hopping matrix in the rotated basis is obtained from
\begin{equation}
\tilde{H}_{lmn}=R^T(A)H_{lmn}R(B).
\end{equation}
Referring to Fig. \ref{fig-cube}, we have $R(A) = R (\theta_r, -\theta_t)$, while $R(B) = R (-\theta_r, \theta_t)$ for hopping to the B atom located along $\hat x$ or $\hat y$ and $R (B) = R (\theta_r, \theta_t)$ along $\hat z$. Thus, for example, $\tilde{H}_{100}  =  R (\theta_r, -\theta_t)^T H_{100}  R (-\theta_r, \theta_t)$. Using their explicit forms already given, the hopping matrices in the rotated basis are readily calculated. They read
\begin{widetext}
{\scriptsize
\begin{equation}
 \tilde{H}_{100} = H_{100}+
                         \frac{1}{\sqrt{2}}\begin{pmatrix}
                                    0 & \hspace*{-0.5cm}V_{\pi}\theta_t  & \hspace*{-0.5cm}-2\theta_tV_{\pi} &\hspace*{-0.25cm} -(3V_{\sigma}+4V_{\pi})\theta_r/\sqrt{2} & \sqrt{3/2}V_{\sigma}\theta_r \\
                                     
                   -\theta_tV_{\pi}& \hspace*{-0.5cm} 0& \hspace*{-0.5cm}-\sqrt{2}V_{\pi}\theta_r&\hspace*{-0.25cm} 0& 0\\
                   
                2\theta_tV_{\pi} & \hspace*{-0.5cm}\sqrt{2}V_{\pi}\theta_r & \hspace*{-1cm}0&\hspace*{-0.25cm}-(3V_{\sigma}+2V_{\pi})\theta_t/2&\sqrt{3}(V_{\sigma}+2V_{\pi})\theta_t/2 \\
                
                (3V_{\sigma}+4V_{\pi})\theta_r/\sqrt{2}& \hspace*{-0.5cm}0 & \hspace*{-0.5cm}(3V_{\sigma}+2V_{\pi})\theta_t/2&\hspace*{-0.25cm} 0&0  \\
                
                   -\sqrt{3/2}V_{\sigma}\theta_r& \hspace*{-0.5cm}0 & \hspace*{-0.5cm}-\sqrt{3}(V_{\sigma}+2V_{\pi})\theta_t/2 &\hspace*{-0.25cm}0&0
\end{pmatrix}, \nonumber
\end{equation}}

{\scriptsize
\begin{equation}
\tilde{H}_{010}  = H_{010}+ \frac{1}{\sqrt{2}}
\begin{pmatrix}
0  &  2\theta_tV_{\pi}   &  \hspace*{-0.6cm}  -\theta_tV_{\pi} 
& \hspace*{-0.05cm}-(3V_{\sigma}+4V_{\pi})\theta_r/\sqrt{2}& -\sqrt{3/2}V_{\sigma}\theta_r\\
                                     
-2\theta_tV_{\pi}&0  &\hspace*{-0.6cm}-\sqrt{2}V_{\pi}\theta_r 
 &\hspace*{-0.05cm}-(3V_{\sigma}+2V_{\pi})\theta_t/2   &-\sqrt{3}(V_{\sigma}+2V_{\pi})\theta_t/2\\
                    
                    \theta_tV_{\pi} &\sqrt{2}V_{\pi}\theta_r&\hspace*{-0.4cm} 0& \hspace*{-0.25cm}0& 0 \\
                    
                    (3V_{\sigma}+4V_{\pi})\theta_r/\sqrt{2} &(3V_{\sigma}+2V_{\pi})\theta_t/2&\hspace*{-0.4cm}0&\hspace*{-0.25cm}0&0  \\
                     \sqrt{3/2}V_{\sigma}\theta_r& \sqrt{3}(V_{\sigma}+2V_{\pi})\theta_t /2&\hspace*{-0.4cm} 0 &\hspace*{-0.25cm} 0 &0
                    
 \end{pmatrix},\nonumber
\end{equation}}
{\scriptsize
\begin{equation}
{\rm and}  \hspace{8 mm}\tilde{H}_{001} =
H_{001}+\frac{1}{\sqrt{2}}\theta_t\begin{pmatrix}
                    0& V_{\pi}&-V_{\pi}& 0 &0\\
                    
                   -V_{\pi}& 0  &0&-V_{\pi}&-\sqrt{3}(V_{\sigma}+V_{\pi})\\
                   
               V_{\pi}&0&0&-V_{\pi}&\sqrt{3}(V_{\sigma}+V_{\pi})\\                 
                  0 &V_{\pi}   &V_{\pi}& 0 & 0\\
                   0& \sqrt{3}(V_{\sigma}+V_{\pi}) &-\sqrt{3}(V_{\sigma}+V_{\pi})& 0 & 0
                   \end{pmatrix}.
\label{eqna8}
\end{equation}}
The TB hopping matrices for other pairs of atoms may be similarly calculated.

{\it TB hopping Integrals} -- For ready reference, we provide the TB hopping integrals between two atoms. The TB hopping integrals $H_{i\alpha, j\beta}$ between the d orbitals located on two sites $i$ and $j$  in the orbital basis $\alpha$ = $xy$, $yz$, $xz$, $x^2-y^2$ and $3z^2-r^2$ are given by\cite{Harrison} 
\begin{equation}
H =H_{\pi}+H_{\sigma}+H_{\delta}, 
\end{equation}
where
{\scriptsize
\[H_{\pi}=V_{\pi}
\begin{pmatrix}
l^2+m^2-4l^2m^2&ln(1-4m^2) & mn(1-4l^2) &2lm(m^2-l^2) &-2\sqrt{3}lmn^2\\
ln(1-4m^2)&m^2+n^2-4m^2n^2 &lm(1-4n^2)&-mnw_{+}&-\sqrt{3}mnw_{0} \\
 mn(1-4l^2)&lm(1-4n^2)&l^2+n^2-4l^2n^2 &nlw_{-} &\sqrt{3}nlw_{0}\\
2lm(m^2-l^2)& -mnw_{+}&nlw_{-}&l^2+m^2-(l^2-m^2)^2 &\sqrt{3}n^2(m^2-l^2) \\
-2\sqrt{3}lmn^2&-\sqrt{3}mnw_{0} & \sqrt{3}nlw_{0}&\sqrt{3}n^2(m^2-l^2)&3n^2(l^2+m^2)
\end{pmatrix},
\]}
{\scriptsize
\[H_{\sigma}=V_{\sigma}\begin{pmatrix}
 3l^2m^2&3lm^2n &3l^2mn &\frac{3}{2}lm(l^2-m^2)&\sqrt{3}lmv\\
3lm^2n&3m^2n^2 &3lmn^2 &\frac{3}{2}mn(l^2-m^2)&  \sqrt{3}mnv\\
3l^2mn&3lmn^2 &3l^2n^2 & \frac{3}{2}nl(l^2-m^2)&\sqrt{3}nlv \\
\frac{3}{2}lm(l^2-m^2)&\frac{3}{2}mn(l^2-m^2) &\frac{3}{2}nl(l^2-m^2) &\frac{3}{4}(l^2-m^2)^2&\frac{\sqrt{3}}{2}(l^2-m^2)w\\
 \sqrt{3}lmv&  \sqrt{3}mnv& \sqrt{3}nlv&\frac{\sqrt{3}}{2}(l^2-m^2)w&v^2
         \end{pmatrix},
\]}
and
{\scriptsize
\begin{equation}
H_{\delta}=V_{\delta}\begin{pmatrix}
(n^2+l^2m^2)&ln(m^2-1) & mn(l^2-1) &\frac{1}{2}lm(l^2-m^2) &\frac{\sqrt{3}}{2}lm(1+n^2)\\
ln(m^2-1)&(l^2+m^2n^2) &lm(n^2-1) &mn[1+\frac{1}{2}(l^2-m^2)] &-\frac{\sqrt{3}}{2}mn(l^2+m^2) \\
mn(l^2-1)&lm(n^2-1)&(m^2+l^2n^2) &-nlu &-\frac{\sqrt{3}}{2}nl(l^2+m^2)\\
\frac{1}{2}lm(l^2-m^2) &mn[1+\frac{1}{2}(l^2-m^2)]  &-nlu  &n^2+\frac{1}{4}(l^2-m^2)^2 &\frac{\sqrt{3}}{4}(1+n^2)(l^2-m^2) \\
\frac{\sqrt{3}}{2}lm(1+n^2)&-\frac{\sqrt{3}}{2}mn(l^2+m^2) & -\frac{\sqrt{3}}{2}mn(l^2+m^2)&\frac{\sqrt{3}}{4}(1+n^2)(l^2-m^2) &\frac{3}{4}(l^2+m^2)^2
\end{pmatrix}.
\end{equation}}
Here, $l, m, $ and $n$ denote the direction cosines of the distance vector ($\vec r_j - \vec r_i$) between the atoms,
$u=[1-\frac{1}{2}(l^2-m^2)]$, 
$v=[n^2-\frac{1}{2}(l^2+m^2)]$,                   $w=[n^2-\frac{1}{2}(l^2+m^2)^2]$, 
$w_{\pm}=[1\pm2(l^2-m^2)] $,                 and $w_{0}=(l^2+m^2-n^2)$.
%
%In the rotated orbital basis, the new hopping integrals $\tilde{H}$ are given by   $ \tilde{H} =R_i^T  H R_j, $ where $R_i$ rotates the orbitals at site $i$.

{\it Rotation of the   L = 1 and L = 2 cubic harmonics in terms of the Euler angles} -- 
The rotation matrix $R_L$ for $L = 1$ cubic harmonics ($x$, $y$, and $z$) or for ordinary vectors is expressed in terms of the  Euler angles as
\begin{equation}
R_1  =\begin{pmatrix}
     \cos\alpha\cos\beta\cos\gamma-\sin\alpha\sin\gamma & \hspace*{0.2cm}-\cos\alpha\cos\beta\sin\gamma-\sin\alpha\cos\gamma & \hspace*{0.2cm}\sin\beta\cos\alpha\\
     \sin\alpha\cos\beta\cos\gamma+\cos\alpha\sin\gamma & \hspace*{0.2cm} -\sin\alpha\cos\beta\sin\gamma+\cos\alpha\cos\gamma & \hspace*{0.2cm}\sin\alpha\sin\beta\\
     -\sin\beta\cos\gamma &  \sin\beta\sin\gamma &  \cos \beta
\end{pmatrix}.
\end{equation}
Sometimes it is useful to make a rotation  of angle $\theta$  about a given axis $\hat u = (l, m, n)$.
In this case, the rotation matrix is given by
%\cos \theta 
\begin{equation}
R_1 ( \theta) =  \begin{pmatrix}
     \cos \theta  + l^2 (1-\cos \theta )  &\hspace*{0.3cm}  lm (1-\cos \theta )-n\sin \theta & \hspace*{0.3cm} ln(1-\cos \theta )+ m\sin \theta \\
        lm(1-\cos \theta )+ n \sin \theta & \cos \theta  +m^2 (1-\cos \theta ) & mn (1-\cos \theta ) - l \sin \theta \\
        ln(1-\cos \theta )- m \sin \theta & mn (1-\cos \theta ) + l \sin \theta & \cos \theta  + n^2 (1-\cos \theta )
\end{pmatrix}.
\end{equation}

For the  $L=2$ cubic harmonics ($xy$, $yz$, $xz$, $x^2-y^2$ and $3z^2-r^2$), the expression for the rotation matrix 
$R_2 (\alpha, \beta, \gamma)$ is
{\scriptsize
\begin{equation}
R_2 =    \frac{1}{2}
 \begin{pmatrix} %{c|c|c|c|c}
\beta^{'2}_+ \cos(2\alpha+2\gamma)/2& \beta^{'}_{+}\sin\beta\cos(2\alpha+\gamma)& \beta^{'}_{+}\sin\beta\sin(2\alpha+\gamma)&\beta^{'2}_{+} \sin(2\alpha+2\gamma)/2&\sqrt{3}\sin^2\beta\\

 -\beta^{'2}_{-}\cos(2\alpha-2\gamma)/2&+\beta^{'}_{-}\sin\beta\cos(2\alpha-\gamma)&-\beta^{'}_{-}\sin\beta\sin(2\alpha-\gamma)&+\beta^{'2}_{-}\sin(2\alpha-2\gamma)/2 &\sin2\alpha\\
& & & &\\
\hline
-\beta^{'}_{+}\sin\beta\cos(\alpha+2\gamma)&\beta^{'}_{+}\beta^{''}_{-}\cos(\alpha+\gamma)&\beta^{'}_{+}\beta^{''}_{-}\sin(\alpha+\gamma)&-\beta^{'}_{+}\sin\beta\sin(\alpha+2\gamma)&\sqrt{3}\sin\alpha\\

-\beta^{'}_{-}\sin\beta\cos(\alpha-2\gamma)&+\beta^{'}_{-}\beta^{''}_{+}\cos(\alpha-\gamma)&-\beta^{'}_{-}\beta^{''}_{+}\sin(\alpha-\gamma)&+\beta^{'}_{-}\sin\beta\sin(\alpha-2\gamma) &\sin2\beta\\ 
& & & &\\
\hline
\beta^{'}_{+}\sin\beta\sin(\alpha+2\gamma)&-\beta^{'}_{+}\beta^{''}_{-}\sin(\alpha+\gamma)&\beta^{'}_{+}\beta^{''}_{-}\cos(\alpha+\gamma)&-\beta^{'}_{+}\sin\beta\cos(\alpha+2\gamma)&\sqrt{3}\cos\alpha\\

+\beta^{'}_{-}\sin\beta\sin(\alpha-2\gamma)&-\beta^{'}_{-}\beta^{''}_{+}\sin(\alpha-\gamma)&-\beta^{'}_{-}\beta^{''}_{+}\cos(\alpha-\gamma)&+\beta^{'}_{-}\sin\beta\cos(\alpha-2\gamma) &\sin2\beta\\ 
& & & &\\

\hline
-\beta^{'2}_{+} \sin(2\alpha+2\gamma)/2&-\beta^{'}_{+}\sin\beta\sin(2\alpha+\gamma)& \beta^{'}_{+}\sin\beta\cos(2\alpha+\gamma)&\beta^{'2}_{+} \cos(2\alpha+2\gamma)/2&\sqrt{3}\sin^2\beta\\

 +\beta^{'2}_{-}\sin(2\alpha-2\gamma)/2&-\beta^{'}_{-}\sin\beta\sin(2\alpha-\gamma)&-\beta^{'}_{-}\sin\beta\cos(2\alpha-\gamma)&+\beta^{'2}_{-}\cos(2\alpha-2\gamma)/2 &\cos2\alpha\\
& & & &\\
\hline
-\sqrt{3}\sin^2\beta\sin2\gamma&\sqrt{3}\sin2\beta\sin\gamma &-\sqrt{3}\sin2\beta\cos\gamma &\sqrt{3}\sin^2\beta\cos2\gamma&2-3\sin^2\beta\\
\end{pmatrix},
\label{R2}
\end{equation}}
where $\beta^\prime_\pm = 1\pm \cos\beta$ and $\beta^{''}_{\pm} = 2 \cos\beta\pm 1$.
\end{widetext}

\section*{Acknowledgement}
We thank Zoran S. Popovi\'{c} for helpful discussions and the U.S. Department of Energy, Office of Basic Energy Sciences, Division of Materials Sciences and Engineering (Grant No. DE-FG02-00ER45818) for financial support. Computational resources were provided by the National Energy Research Scientific Computing Center, a user facility also supported by the U.S. Department of Energy.


\begin{thebibliography}{99}

\bibitem{shi_PRB_2009} Y. G. Shi, Y. F. Guo, S. Yu, M. Arai, A. A. Belik, A. Sato, K. Yamaura, E. Takayama-Muromachi, H. F. Tian, H. X. Yang, J. Q. Li, T. Varga, J. F. Mitchell, and S. Okamoto, Phys. Rev. B {\bf 80}, 161104(R) (2009).

\bibitem{calder_PRL_2012} S. Calder, V. O. Garlea, D. F. McMorrow, M. D. Lumsden, M. B. Stone, J. C. Lang, J.-W. Kim, J. A. Schlueter, Y. G. Shi, K. Yamaura, Y. S. Sun, Y. Tsujimoto, and A. D. Christianson, Phys. Rev. Lett. {\bf 108}, 257209 (2012).

\bibitem{calder_PRB_2017} S. Calder, J. G. Vale, N. Bogdanov, C. Donnerer, D. Pincini, M. Moretti Sala, X. Liu, M. H. Upton, D. Casa, Y. G. Shi, Y. Tsujimoto, K. Yamaura, J. P. Hill, J. van den Brink, D. F. McMorrow, and A. D. Christianson, Phys. Rev. B {\bf 95}, 020413(R) (2017).

\bibitem{kermarrec_PRB_2015} E. Kermarrec, C. A. Marjerrison, C. M. Thompson, D. D. Maharaj, K. Levin, S. Kroeker, G. E. Granroth, R. Flacau, Z. Yamani, J. E. Greedan, and B. D. Gaulin, Phys. Rev. B {\bf 91}, 075133 (2015).

\bibitem{taylor_PRB_2016} A. E. Taylor, R. Morrow, R. S. Fishman, S. Calder, A. I. Kolesnikov, M. D. Lumsden, P. M. Woodward, and A. D. Christianson, Phys. Rev. B {\bf 93}, 220408(R) (2016).

\bibitem{taylor_PRL_2017} A. E. Taylor, S. Calder, R. Morrow, H. L. Feng, M. H. Upton, M. D. Lumsden, K. Yamaura, P. M. Woodward, and A. D. Christianson, Phys. Rev. Lett. {\bf 118}, 207202 (2017).

\bibitem{du_PRB_2012} Y. Du, X. Wan, L. Sheng, J. Dong, and S. Y. Savrasov, Phys. Rev. B {\bf 85}, 174424 (2012).

\bibitem{jung_PRB_2013} M.-C. Jung, Y.-J. Song, K.-W. Lee, and W. E. Pickett, Phys. Rev. B {\bf 87}, 115119 (2013).

\bibitem{zahid_JPCS_2015} Z. Ali, A. Sattar, S. J. Asadabadi, and I. Ahmad, J. Phys. \& Chem. Solids {\bf 86}, 114 (2015).

\bibitem{morrow_CM_2016} R. Morrow, K. Samanta, T. Saha Dasgupta, J. Xiong, J. W. Freeland,.D. Haskel, and P. M. Woodward, Chem. Mater. {\bf 28}, 3666 (2016). 

\bibitem{kim_PRB_2016} B. Kim, P. Liu, Z. Erg\"{o}nenc, A. Toschi, S. Khmelevskyi, and C. Franchini, Phys. Rev. B {\bf 94}, 241113(R) (2016).

\bibitem{vecchio_SCI_REP_2013} I. Lo Vecchio, A. Perucchi, P. Di Pietro, O. Limaj, U. Schade, Y. Sun, M. Arai, K. Yamaura, and S. Lupi, Sci. Rep. {\bf 3}, 2990 (2013).

\bibitem{vale_arxiv_2017} J. G. Vale, S. Calder, C. Donnerer, D. Pincini, Y. G. Shi, Y. Tsujimoto, K. Yamaura, M. Moretti Sala, J. van den Brink, A. D. Christianson, and D. F. McMorrow, arXiv:1707.05551 (2017).

%============ DFT references  =========================

\bibitem{Methfessel} M. Methfessel, M. van Schilfgaarde, and R. A.  Casali, A Full-Potential LMTO Method Based on Smooth Hankel Functions, Electronic Structure and Physical Properties of Solids. The Use of the LMTO Method, Lecture Notes in Physics \textbf{535}, 114 (2000).

\bibitem{Kotani10} T. Kotani and M. van Schilfgaarde, Fusion of the LAPW and LMTO methods: The augmented plane wave plus muffin-tin orbital method, Phys. Rev. B \textbf{81}, 125117 (2010).

\bibitem{lmsuite} https://www.questaal.org

\bibitem{vonBarthHedin} U. von Barth and L. Hedin, A local exchange-correlation potential for the spin polarized case, Journal of Physics C: Solid State Physics \textbf{5}, 1629 (1972).

\bibitem{iridate_paper} S. Mohapatra, J. van den Brink, and A. Singh, Phys. Rev. B {\bf 95}, 094435 (2017).

\bibitem{fang_PRB_2001} Z. Fang and K. Terakura, Phy. Rev. B {\bf 64}, 020509(R) (2001). 

\bibitem{osmate_new} A. Singh, S. Mohapatra, C. Bhandari, and S. Satpathy, arXiv:1802.01449 (2018).

%\bibitem{Kohn-Sham} W. Kohn and L. J. Sham, Self-Consistent Equations Including Exchange and Correlation Effects, Phys. Rev. \textbf{140}, A1133 (1965).

%\bibitem{KimPRB016}K. Bongjae {\sl et al.}, Lifshitz transition driven by spin fluctuations and spin-orbit renormalization in ${\rm NaOsO_3}$, Phys. Rev. B \textbf{94} (2016).

%\bibitem{calder_PRL_2012} S. Calder {\sl et al.}, Magnetically Driven Metal-Insulator Transition in ${\rm NaOsO_3}$, Phys. Rev. Lett. \textbf{108}, 257209 (2012).

%\bibitem{Bradley} C. J. Bradley and A. P. Cracknell, {\it The mathematical theory of symmetry in solids}, Oxford University Press, Oxford (1972).

%\bibitem{du_PRB_2012} Y. Du, X. Wan, L. Sheng, J. Dong, and S. Y. Savrasov, Electronic structure and magnetic properties of NaOsO${}_{3}$, Phys. Rev. B \textbf{85}, 174424 (2012).

%\bibitem{jung_PRB_2013} M. C. Jung {\sl et al.}, Structural and correlation effects in the itinerant insulating antiferromagnetic perovskite NaOsO${}_{3}$, Phys. Rev. B \textbf{87}, 115119 (2013).
 
% appendix reference
 
\bibitem{Tinkham} M. Tinkham, {\it Group Theory and Quantum Mechanics}, McGraw Hill, New York (1964).
 
\bibitem{Harrison} W. A. Harrison, {\it Electronic Structure and the Properties of Solids}, Dover, New York (1989).

\end{thebibliography}
\end{document}